\pdfoutput=1
   \documentclass[fleqn,usenatbib,usedcolumn]{mnras}
   \usepackage[british]{babel}             
   \usepackage{newtxtext}                  
   \usepackage[slantedGreek]{newtxmath}    
   \let\la=\lesssim 

\usepackage[T1]{fontenc}
\usepackage{ae,aecompl}

\usepackage{amsmath}	
\usepackage{amssymb}	
\usepackage{graphicx,color}
\usepackage{pdflscape}
\usepackage{times,tikz}
\usepackage{float}

\def\hi{\mbox{H\sc{i}}}
\def\kms{km~s$^{-1}$}

\def\msun{M$_{\odot}$}
\def\atlas{ATLAS$^{\rm 3D}$}
\def\arcsec{$^{\prime \prime}$}
\definecolor{Mygrey}{gray}{0.75}

\newcommand{\ltsimeq}{\raisebox{-0.6ex}{$\,\stackrel{\raisebox{-.2ex}{$\textstyle <$}}{\sim}\,$}}

\newcommand{\farc}{\mbox{\ensuremath{.\!\!^{\prime\prime}}}}

\newcommand{\coordsec}{\mbox{\ensuremath{.\!\!^{\rm s}}}}

\mathchardef\mhyphen="2D

\usepackage[compact]{titlesec}

   \hypersetup{pdfauthor={T. A. Davis},
               pdftitle={WISDOM Project -- III: Molecular gas measurement of the supermassive black hole mass in the barred lenticular galaxy NGC4429},
               pdfkeywords={galaxies: individual: NGC 4429 -- galaxies: kinematics and dynamics -- galaxies: nuclei -- galaxies: ISM -- galaxies: elliptical and lenticular, cD},
               bookmarksnumbered=true}
\usepackage{etoolbox}
\makeatletter
\patchcmd\@combinedblfloats{\box\@outputbox}{\unvbox\@outputbox}{}{%
  \errmessage{\noexpand\@combinedblfloats could not be patched}%
}
\makeatother

   \volume{{\rm in press}}
   
\titlespacing{\section}{0pt}{*2}{*1}

\title[WISDOM: The SMBH in NGC4429]{WISDOM Project -- III: Molecular gas measurement of the supermassive black hole mass in the barred lenticular galaxy NGC4429} 
\author[Timothy A. Davis et al.]{\parbox{\textwidth}{Timothy A. Davis,$^{1}$\thanks{E-mail: \texttt{DavisT@cardiff.ac.uk}} Martin Bureau,$^{2}$ Kyoko Onishi,$^{3,4,5}$ Freeke van de Voort,$^{6,7}$
Michele Cappellari,$^{2}$
Satoru Iguchi,$^{3,4}$
Lijie Liu,$^{2}$
Eve V. North,$^{1}$
Marc Sarzi$^{8}$
and Mark D. Smith$^{2}$
}
\vspace{0.4cm}\\
\parbox{\textwidth}{$^{1}$School of Physics \&\ Astronomy, Cardiff University, Queens Buildings, The Parade, Cardiff, CF24 3AA, UK\\
$^{2}$Sub-department of Astrophysics, Department of Physics, University of Oxford, Denys Wilkinson Building, Keble Road, Oxford OX1 3RH, UK\\
$^{3}$Department of Astronomical Science, SOKENDAI (The Graduate University of Advanced Studies), Mitaka, Tokyo 181-8588, Japan\\
$^{4}$National Astronomical Observatory of Japan, Mitaka, Tokyo 181-8588, Japan\\
$^{5}$Research Center for Space and Cosmic Evolution, Ehime University, 2-5 Bunkyo-cho, Matsuyama, Ehime 790-8577, Japan\\
$^{6}$Heidelberg Institute for Theoretical Studies, Schloss-Wolfsbrunnenweg 35, 69118, Heidelberg, Germany\\
$^{7}$Astronomy Department, Yale University, PO Box 208101, New Haven, CT 06520-8101, USA\\
$^{8}$Centre for Astrophysics Research, University of Hertfordshire, Hatfield, Hertfordshire, AL1 9AB, UK
}}
\begin{document}
\date{Accepted 2017 October 4. Received 2017 October 3; in original form 2017 May 10.}

\pagerange{\pageref{firstpage}--\pageref{lastpage}} \pubyear{2015}

\maketitle

\label{firstpage}

\begin{abstract}
As part of the mm-Wave Interferometric Survey of Dark Object Masses (WISDOM) project we present an estimate of the mass of the supermassive black hole (SMBH) in the nearby fast-rotating early-type galaxy NGC4429, that is barred and has a boxy/peanut-shaped bulge. This estimate is based on Atacama Large Millimeter/submillimeter Array (ALMA) cycle-2 observations of the $^{12}$CO(3--2) emission line with a linear resolution of $\approx$13 pc (0\farc18\,$\times$\,0\farc14). NGC4429 has a relaxed, flocculent nuclear disc of molecular gas that is truncated at small radii, likely due to the combined effects of gas stability and tidal shear. The warm/dense $^{12}$CO(3-2) emitting gas is confined to the inner parts of this disc, likely again because the gas becomes more stable at larger radii, preventing star formation. The gas disc has a low velocity dispersion of 2.2$^{+0.68}_{-0.65}$ \kms. Despite the inner truncation of the gas disc, we are able to model the kinematics of the gas and estimate a mass of (1.5$\pm0.1^{+0.15}_{-0.35}$)~$\times$10$^8$~\msun\ for the SMBH in NGC4429 (where the quoted uncertainties reflect the random and systematic uncertainties, respectively), consistent with a previous upper limit set using ionised gas kinematics.  We confirm that the $V$-band mass-to-light ratio changes by $\approx$30\% within the inner 400\,pc of NGC4429, as suggested by other authors. This SMBH mass measurement based on molecular gas kinematics, the sixth presented in the literature, once again demonstrates the power of ALMA to constrain SMBH masses.
\end{abstract} 

\renewcommand{\thefootnote}{\ensuremath{^\arabic{footnote}}}
\setcounter{footnote}{0}

\section{INTRODUCTION}
\noindent

The physical processes involved in the evolution of galaxies act on scales that span many orders of magnitude in both time and space. From the effects of large-scale structure to those of small-scale turbulence, and from gas accretion/halo shocks to stellar feedback, understanding how these processes couple with each other across these spatial and temporal dimensions is one of the key challenges in contemporary astrophysics. 

Being physically small, but gravitationally important, and with the potential to power jets and winds that can affect the evolution of galaxies on the largest scales \citep[e.g.][]{1998A&A...331L...1S,2008ApJ...676...33D}, supermassive black holes (SMBHs) provide one of the most acute problems in this galaxy evolution framework.
Key observations that we need to understand if we wish to discern the role of black holes in galaxy formation include the strong correlations between galaxy properties and the masses of their SMBHs \citep[e.g.][]{1998AJ....115.2285M,2000ApJ...539L..13G,2001ApJ...563L..11G,2009ApJ...698..198G,2013ApJ...764..184M,2016ApJ...831..134V}.

Various methods exist to dynamically measure the masses of SMBHs, from both the kinematics of stars \citep[e.g.][]{1988ApJ...324..701D,1998AJ....115.2285M,2003ApJ...583...92G,2002MNRAS.335..517V,2009MNRAS.394..660C} and gas
(e.g. \citealt{1995Natur.373..127M,1996ApJ...470..444F,2001ApJ...550...65S,2007ApJ...671.1329N,2010ApJ...721...26G}). Measurements made using these techniques have led to our current empirical SMBH -- galaxy relations \cite[e.g.][]{2013ARA&A..51..511K,2013ApJ...764..184M,2016ApJ...831..134V}. 

Over the last few years, as ALMA (Atacama Large Millimeter/submillimeter Array) and its precursors grew in power, molecular gas kinematics has begun to be used to estimate SMBH masses. 
The Virgo cluster fast-rotating early-type galaxy (ETG) NGC4526 was the first object to have its SMBH mass measured using this method \citep{2013Natur.494..328D}, and measurements in NGC1097 \citep{2015ApJ...806...39O} and NGC1332 \citep{2016ApJ...822L..28B,2016ApJ...823...51B} soon followed.
\cite{2014MNRAS.443..911D} presented a figure of merit specific to this technique and showed that this method should allow SMBH mass measurements in thousands of galaxies across the universe, while \cite{2014ApJ...791L..41H} discussed the prospects of its use in lensed galaxies at very high redshifts.

 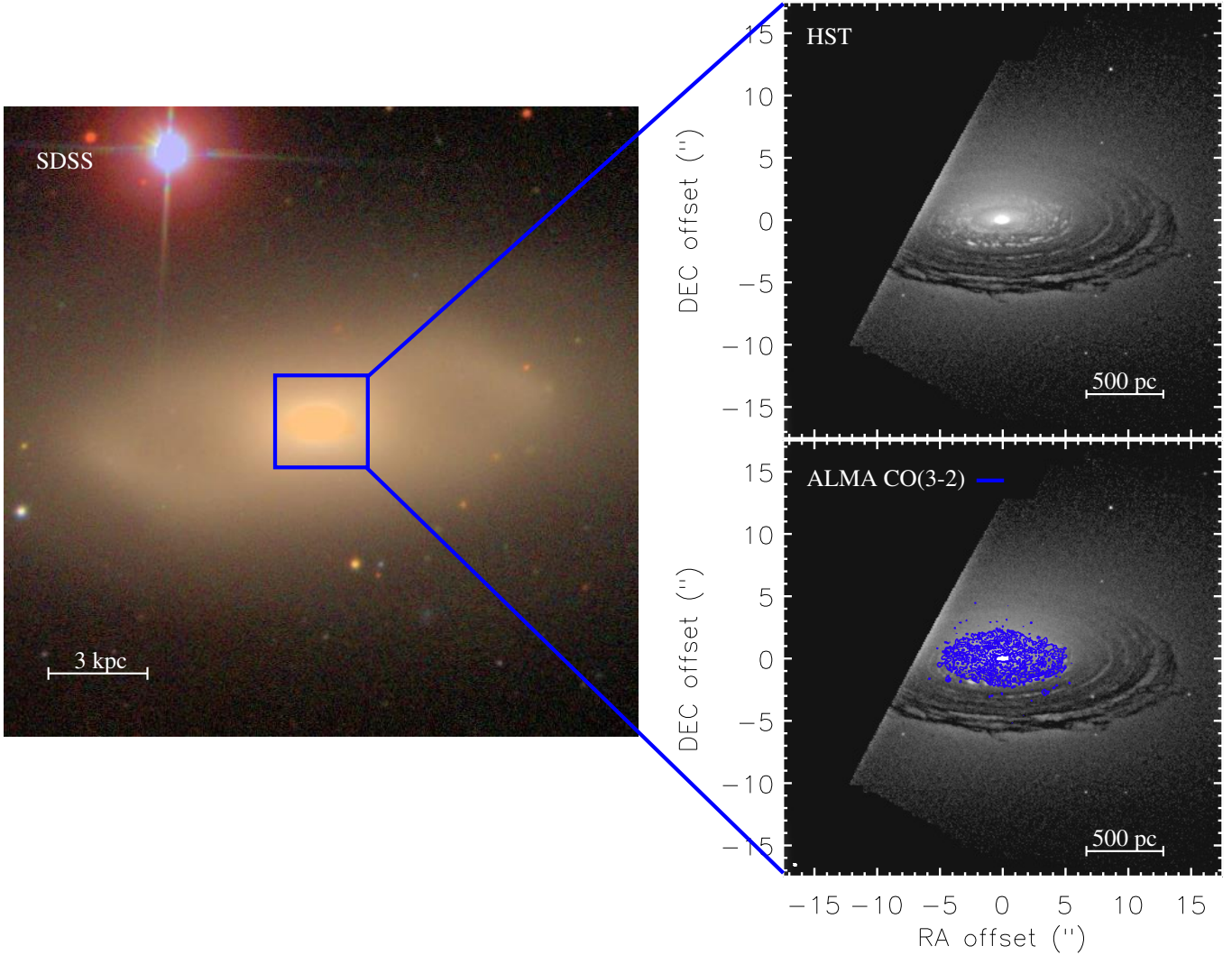
\begin{figure*}\begin{center}
\begin{tikzpicture}
    \node[anchor=south west,inner sep=0] (image) at (0,0) { 
    	\begin{minipage}[b]{0.52\textwidth}
\begin{tikzpicture}
    \node[anchor=south west,inner sep=0] (image) at (0,0) {\includegraphics[height=9.25cm,angle=0,clip,trim=0cm 0cm 0cm 0.0cm]{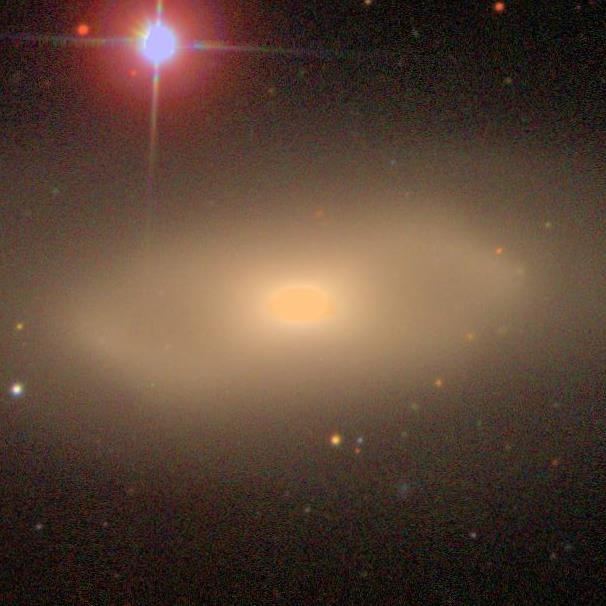}};
    	\begin{scope}[x={(image.south east)},y={(image.north west)}]
        \draw[blue,ultra thick] (0.427,0.427) rectangle (0.573,0.573);
         \node[text=white] at (0.05,0.9) {\large SDSS};
        	\draw[white,thick] (0.07,0.1) -- (0.226,0.1);	
	\draw[white,thick] (0.07,0.09) -- (0.07,0.11);	
	\draw[white,thick] (0.226,0.09) -- (0.226,0.11);	
	 \node[text=white] at (0.11,0.10) {\large 3 kpc};
           \end{scope}
\end{tikzpicture}\vspace{3.2cm}
	\end{minipage}\hspace{0.5cm}
	\begin{minipage}[b]{0.45\textwidth}
	\begin{tikzpicture}
    \node[anchor=south west,inner sep=0] (image) at (0,0) {\includegraphics[width=8cm,angle=0,clip,trim=0cm 2.9cm 0cm 0.0cm]{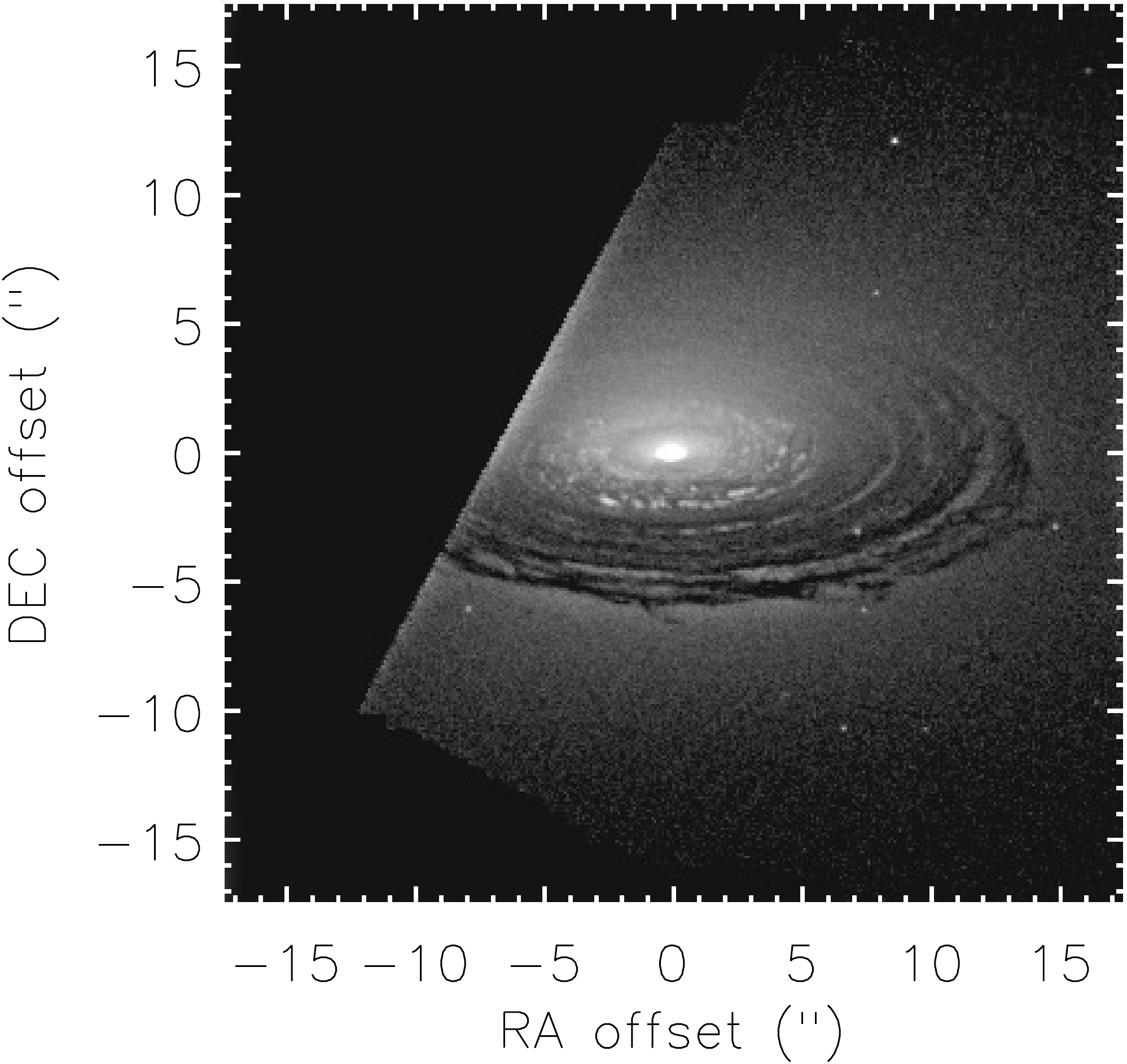}};
     	\begin{scope}[x={(image.south east)},y={(image.north west)}]
 	\draw[white,thick] (0.75,0.1) -- (0.89,0.1);	
	\draw[white,thick] (0.75,0.09) -- (0.75,0.11);	
	\draw[white,thick] (0.89,0.09) -- (0.89,0.11);	
	 \node[text=white] at (0.76,0.1) {\large 500 pc};
           \node[text=white] at (0.24,0.9) {\large HST};
           \end{scope}
\end{tikzpicture}
	\begin{tikzpicture}
    \node[anchor=south west,inner sep=0] (image) at (0,0) {
    \includegraphics[width=8cm,angle=0,clip,trim=0cm 0cm 0cm 0.0cm]{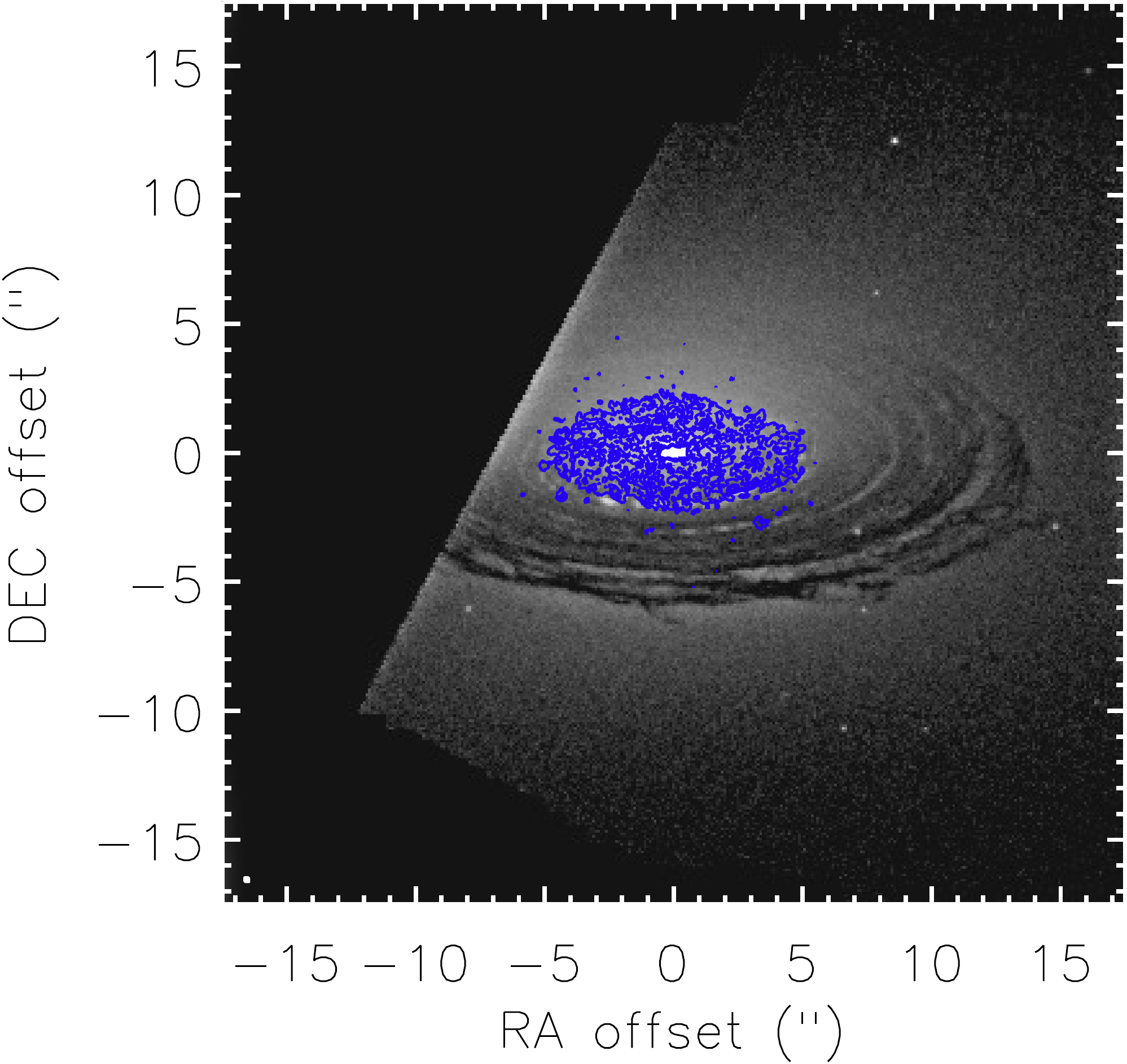}
    };
     	\begin{scope}[x={(image.south east)},y={(image.north west)}]
	\draw[white,thick] (0.75,0.2) -- (0.89,0.2);	
	\draw[white,thick] (0.75,0.19) -- (0.75,0.21);	
	\draw[white,thick] (0.89,0.19) -- (0.89,0.21);	
	 \node[text=white] at (0.76,0.2) {\large 500 pc};
           \node[text=white] at (0.24,0.9) {\large ALMA CO(3-2)};
           \draw[blue,ultra thick] (0.55,0.92) -- (0.6,0.92);	
           \end{scope}
\end{tikzpicture}
	\end{minipage}
	};
    	\begin{scope}[x={(image.south east)},y={(image.north west)}]
        	\draw[blue,ultra thick] (0.298,0.605) -- (0.64,0.997);	
	\draw[blue,ultra thick] (0.298,0.51) -- (0.64,0.085);
           \end{scope}
\end{tikzpicture}
\caption{\textit{Left panel:} SDSS three-colour ($gri$) image of NGC4429, 4$^\prime\times$4$^\prime$ (19.2 kpc $\times$ 19.2 kpc) in size. \textit{Right panel, top:} Unsharp-masked \textit{HST} Wide-Field Planetary Camera 2 (WFPC2) F606W image of a 2.8 kpc $\times$ 2.8 kpc region around the nucleus (indicated in blue in the left panel), revealing a clear central dust disc. \textit{Right panel, bottom:} As above, but overlaid with blue $^{12}$CO(3-2) integrated intensity contours from our ALMA observations. The synthesised beam is shown as a (very small) white ellipse at the bottom left of the panel (0\farc18\,$\times$\,0\farc14 or 14\,$\times$\,11 pc$^2$). The molecular gas disc coincides with the inner part of the dust disc and exhibits a central hole.}
\label{gal_overview}
 \end{center}
 \end{figure*}

This paper is the third in the mm-{W}ave {I}nterferometric {S}urvey of {D}ark {O}bject {M}asses (WISDOM) project. This survey builds on some of these small pilot projects (\citealt{2013Natur.494..328D,2015ApJ...806...39O}), and it aims to benchmark and test the molecular gas dynamics method, develop tools and best practice, and exploit the growing power of ALMA to better populate and thus constrain SMBH -- galaxy scaling relations. 
The first paper in this series \citep{2017arXiv170305247O} discussed in detail the tools and fitting procedures developed so far, and presented a mass measurement in the nearby fast-rotating early-type galaxy NGC 3665 using CARMA (Combined Array for Research in Millimeter-wave Astronomy). The second paper \citep{2017arXiv170305248D} presented an ALMA measurement of the SMBH in the dynamically cold molecular disc of the elliptical galaxy NGC4697.

 \begin{figure*}\hspace{-3cm}
\begin{minipage}{10cm}
\includegraphics[width=9.9cm,angle=0,clip,trim=0cm 0cm 0cm 0.0cm]{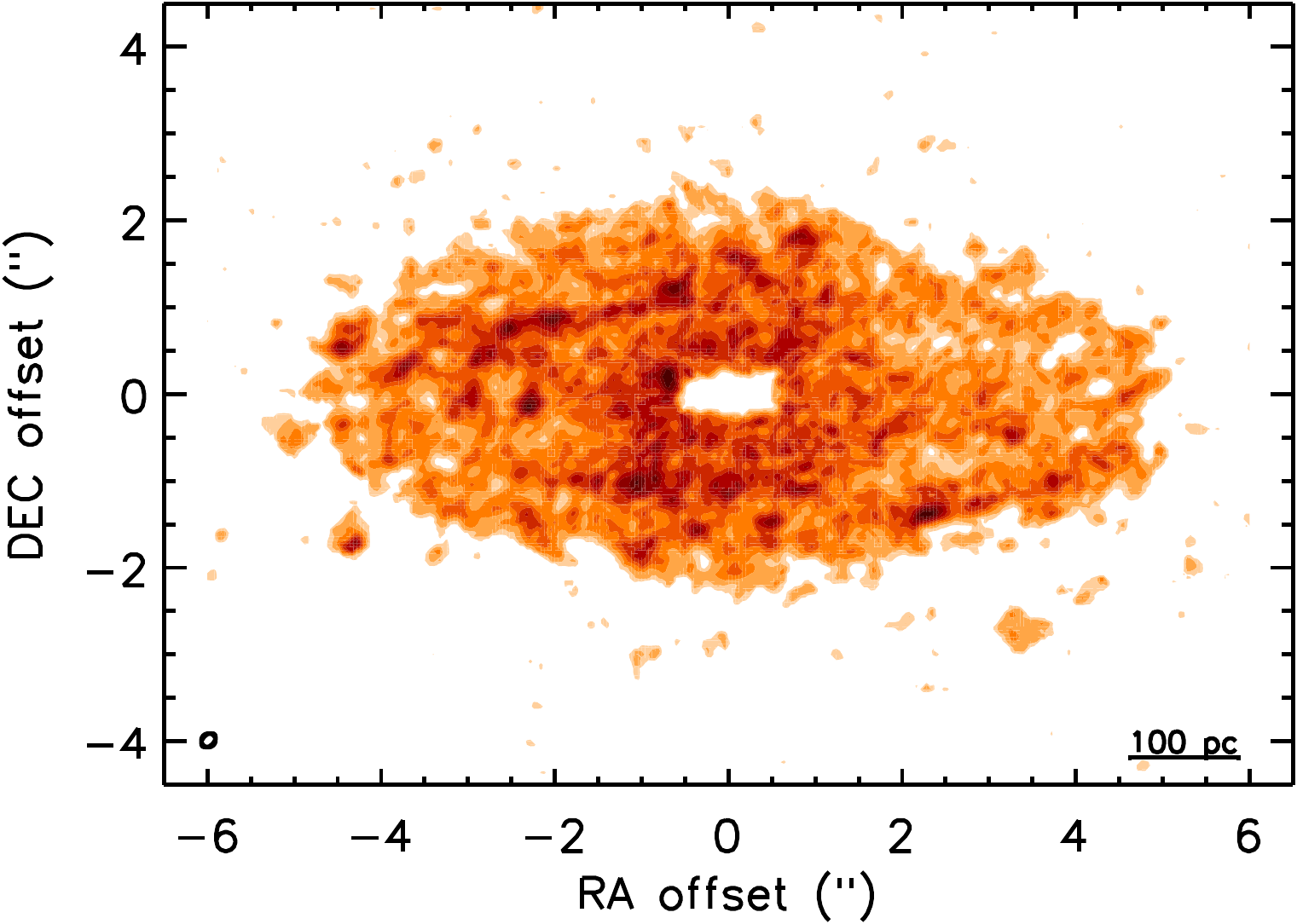}
\end{minipage}\hspace{0.2cm}\begin{minipage}{0.28\textwidth}\includegraphics[width=7.5cm,angle=0,clip,trim=0cm 1.5cm 0cm 0.0cm]{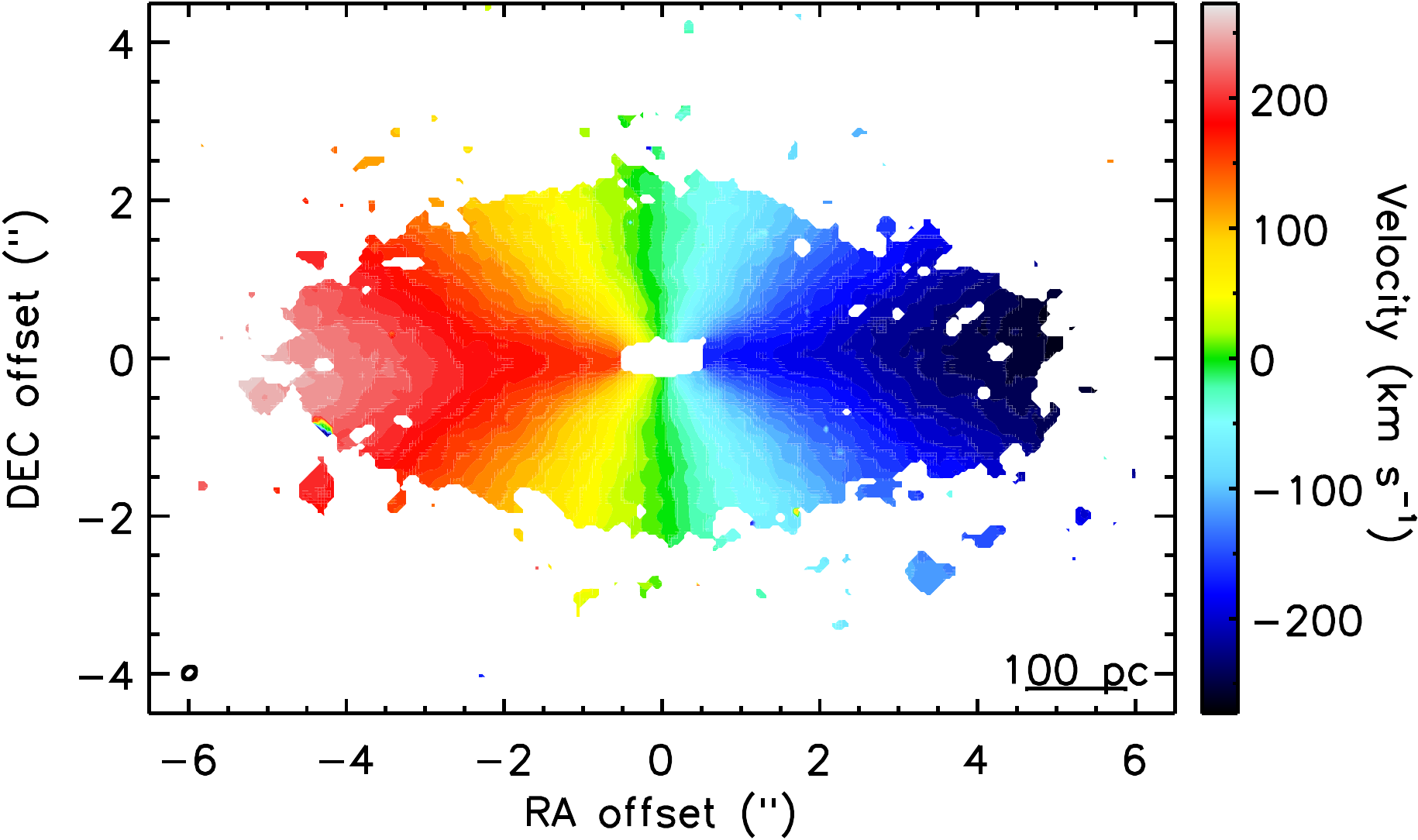}
\includegraphics[width=7.5cm,angle=0,clip,trim=0cm 0cm 0cm 0.0cm]{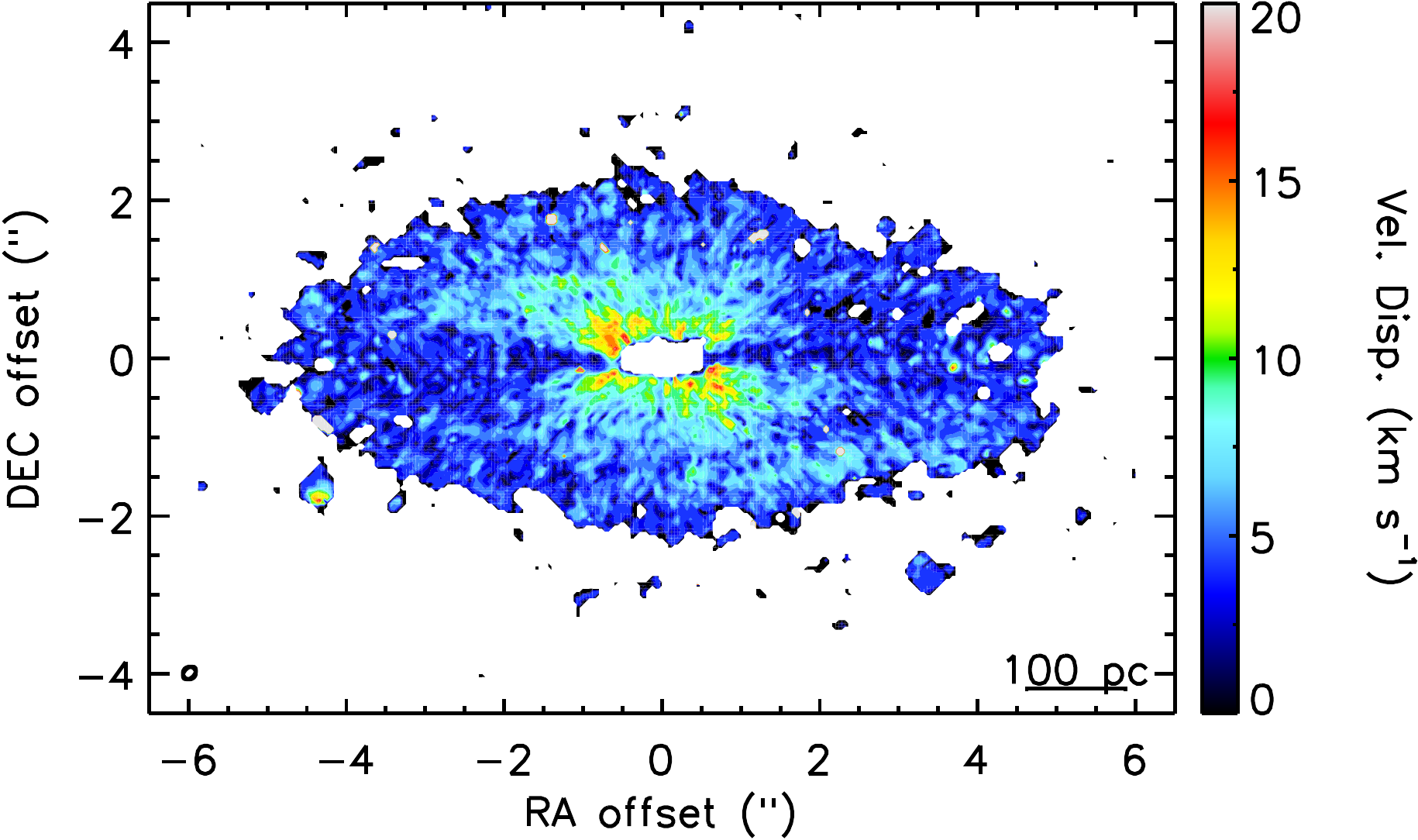}
\end{minipage}
\begin{center}
\caption{Moments of the detected $^{12}$CO(3-2) emission in NGC4429. \textit{Left:} $^{12}$CO(3-2) zeroth moment (integrated intensity) map, created using the masked-moment technique described in Section 3.1.  \textit{Right:} First and second moment (velocity and velocity dispersion) maps of the $^{12}$CO(3-2) emission, created using the same mask. The synthesised beam (0\farc18\,$\times$\,0\farc14 or 14\,$\times$\,11 pc$^2$) is shown as a black ellipse at the bottom left of each panel. }
\label{comomentfig}
 \end{center}
 \end{figure*}

In this work we present ALMA cycle-2 observations of the molecular gas disc in the centre of the fast-rotating Virgo cluster barred lenticular galaxy NGC4429 (see Fig. \ref{gal_overview}), and we use these to estimate the SMBH mass. 
In Section \ref{target} of this paper we describe our target. In Section \ref{data} we present our ALMA observations and the derived data products. In Section \ref{method} we describe our dynamical modelling method. In Section \ref{discuss} we discuss our results before finally concluding in Section \ref{conclude}. Throughout this paper we assume a distance of 16.5$\pm$1.6 Mpc for NGC4429 \citep{2011MNRAS.413..813C}, as derived from the surface brightness fluctuation measurements of \cite{Tonry:2001ei}. At this distance one arcsecond corresponds to a physical scale of $\approx$80 pc.

 \section{\uppercase{Target}}
 \label{target}
\noindent  NGC4429 is a lenticular galaxy with a boxy/peanut-shaped bulge \citep{2013MNRAS.431.3060E}, that lies within the Virgo cluster. Optical integral-field observations as part of the \atlas\ project \citep{2011MNRAS.413..813C} revealed that NGC4429 has a total stellar mass of 1.5$\times$10$^{11}$~\msun, a luminosity-weighted stellar velocity dispersion within one effective radius of  $\sigma_{\rm e}$= 177~\kms\ \citep{2013MNRAS.432.1709C}, and is a fast rotator \citep[$\lambda_{\mathrm{R_e}}$=0.4; ][]{2011MNRAS.414..888E}.

 \textit{Hubble Space Telescope} (\textit{HST}) imaging (Figure \ref{gal_overview}) shows that NGC4429 has a nuclear disc of dust visible in extinction against the stellar continuum. \cite{2011MNRAS.414..940Y} detected (1.1$\pm$0.08)\,$\times\,10^8$\,\msun\ of molecular gas via $^{12}$CO(1-0) single-dish observations, and \cite{2013MNRAS.432.1796A} mapped this gas with CARMA. The $^{12}$CO(1-0) emitting gas lies co-incident with the dust disc and regularly rotates in the galaxy mid-plane \citep{2011MNRAS.417..882D,2013MNRAS.429..534D}. The star formation rate (SFR) within this molecular disc has been estimated at 0.1~\msun\ yr$^{-1}$ by \cite{2014MNRAS.444.3427D} using mid-infrared and far-ultraviolet emission. 

There is unresolved (sub-arcsecond) radio continuum emission from the central regions of NGC4429, suggesting it harbours a low-luminosity active galactic nucleus (AGN; \citealt{2016MNRAS.458.2221N}). Optical spectroscopy also reveals signatures of nuclear activity in NGC4429, including a broad-line region seen in H$\alpha$ \citep{2011PASP..123..514B,2015ApJ...814..149C}.
The mass of the SMBH has not been sucessfully measured to date, but \cite{2009ApJ...692..856B} did set an upper limit of 1.8~$\times$~10$^8$~\msun.
This upper limit is consistent with the prediction of the M$_{\rm BH}$--$\sigma_*$ relation of \cite{2013ApJ...764..184M}, that suggests an SMBH mass of $\approx1.0\,\times$~10$^8$~\msun.

 \section{\uppercase{ALMA data}}
 \label{data}

 The  $^{12}$CO(3--2) line in NGC4429 was observed with ALMA on both the 26th and 27th of June 2015 as part of the WISDOM project (programme 2013.1.00493.S). The total integration time on source was 1.17 hours, split equally between the two tracks. Forty-two of ALMA's 12m antennas were used, arranged such that the data are sensitive to emission on scales up to 11\farc2.
An 1850 MHz correlator window was placed over the $^{12}$CO(3--2) line, yielding a continuous velocity coverage of $\approx$1600 \kms\ with a raw velocity resolution of $\approx$0.4 \kms, sufficient to properly cover and sample the line. Three additional 2 GHz wide low-resolution correlator windows were simultaneously used to detect continuum emission.

The raw ALMA data were calibrated using the standard ALMA pipeline, as provided by the ALMA regional centre staff. Additional flagging was carried out where necessary to improve the data quality. Amplitude and bandpass calibration were performed using the quasars 3C273 and J1229+0203, respectively. The atmospheric phase offsets present in the data were determined using J1229+0203 as a phase calibrator.  

We then used the \texttt{Common Astronomy Software Applications} {(\tt CASA)} package to combine and image the visibility files of the two tracks, producing a three-dimensional RA-Dec-velocity data cube (with velocities determined with respect to the rest frequency of the $^{12}$CO(3-2) line). In this work we primarily use data with a channel width of 10\,\kms, but in Section \ref{uncertainties} we re-image the calibrated visibilities with a channel width of 2\,\kms. In both cases, pixels of 0\farc05 were chosen as a compromise between spatial sampling and resolution, resulting in approximately 3.5 pixels across the synthesised beam.  

The data presented here were produced using Briggs weighting with a robust parameter of 0.5, yielding a synthesised beam of 0\farc18\,$\times$\,0\farc14 at a position angle of 311$^{\circ}$ (a physical resolution of 14\,$\times$\,11 pc$^2$). 
Continuum emission was detected, measured over the full line-free bandwidth, and then subtracted from the data in the $uv$ plane using the {\tt CASA} task {\tt uvcontsub}. The achieved continuum root-mean square (RMS) noise (in both reductions) is 36 $\mu$Jy. The continuum-subtracted dirty cubes were cleaned in regions of source emission (identified interactively) to a threshold equal to the RMS noise of the dirty channels.  The clean components were then added back and re-convolved using a Gaussian beam of full-width-at-half-maximum (FWHM) equal to that of the dirty beam.  This produced the final, reduced, and fully calibrated $^{12}$CO(3--2) data cubes of NGC4429, with RMS noise levels of 0.61 mJy beam$^{-1}$ in each 10 \kms\ channel (and 1.34 mJy beam$^{-1}$ in 2 \kms\ channels).

  \begin{figure} \begin{center}
\includegraphics[width=0.5\textwidth,angle=0,clip,trim=0cm 0cm 0cm 0.0cm]{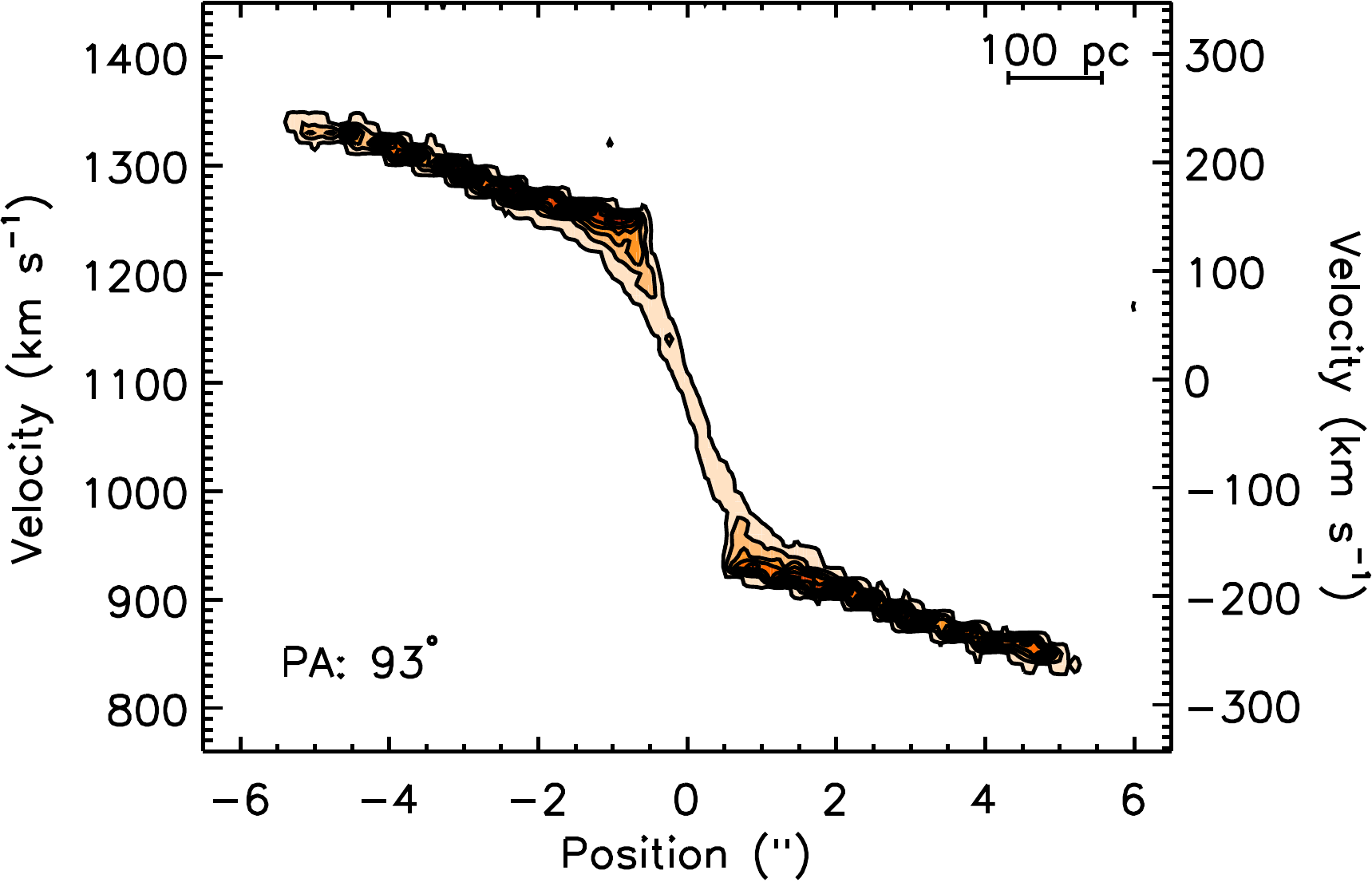}
\caption{Position-velocity diagram of the $^{12}$CO(3-2) emission in NGC4429, extracted along the kinematic major axis with a slit 5 pixels in width. We do not show the beamsize (0\farc18\,$\times$\,0\farc14) or velocity channel width (10 \kms) explicitly in this plot, as they are very small compared to the ranges plotted. Clear channelisation is present in the PVD, suggesting the velocity dispersion in the gas is small compared to our channel width.}
\label{pvdplot}
 \end{center}
 \end{figure}

 \subsection{Line emission}
 \label{lineemission}
 
For each of the two velocity binnings used, the clean, fully-calibrated data cube produced as described above was used to create our final data products. 
A zeroth moment (integrated intensity) map, first moment (mean velocity) map, and second moment (velocity dispersion) map of the detected line emission were created using a masked moment technique. As in our previous works, a copy of the clean data cube was Gaussian-smoothed spatially (with a FWHM equal to that of the synthesised beam), and then Hanning-smoothed using a 3 channel window in velocity. A three-dimensional mask was then defined by selecting all pixels above a fixed flux threshold of 1.5 times the RMS noise, selected to recover as much flux as possible in the moment maps while minimising the noise.  The moment maps were then created using the original un-smoothed cubes within the masked regions only, without any threshold.
The moments of the 10~\kms\ channel$^{-1}$ cube are presented in Figure \ref{comomentfig}.

We clearly detect a flocculent disc of molecular gas in NGC4429, with a radius of $\approx$400 pc. This disc has a central hole of $\approx$40 pc in radius. The gas is regularly rotating and lies only in the inner part of the dust disc visible in \textit{HST} images (see Fig. \ref{gal_overview}). We discuss the morphology of the gas further in Section \ref{discuss}.

A major-axis position-velocity diagram (PVD; taken with a position angle of 93$^{\circ}$, as determined below, and a width of 5 pixels) was extracted from the same data cube and is shown in Figure \ref{pvdplot}.
The channel maps are presented in Section \ref{channelmaps}.
The velocity dispersion in this molecular gas disc seems very low, with clear channelisation present in the PVD (Figure \ref{pvdplot}), similar to that seen in NGC4697 \citep{2017arXiv170305248D}. This suggests that the velocity dispersion is small compared to our channel width of 10 \kms. 
This is discussed further in Sections \ref{veldisp1} and \ref{veldispdiscuss}.

\begin{figure} \begin{center}
\includegraphics[width=0.48\textwidth,angle=0,clip,trim=0cm 0cm 0cm 0.0cm]{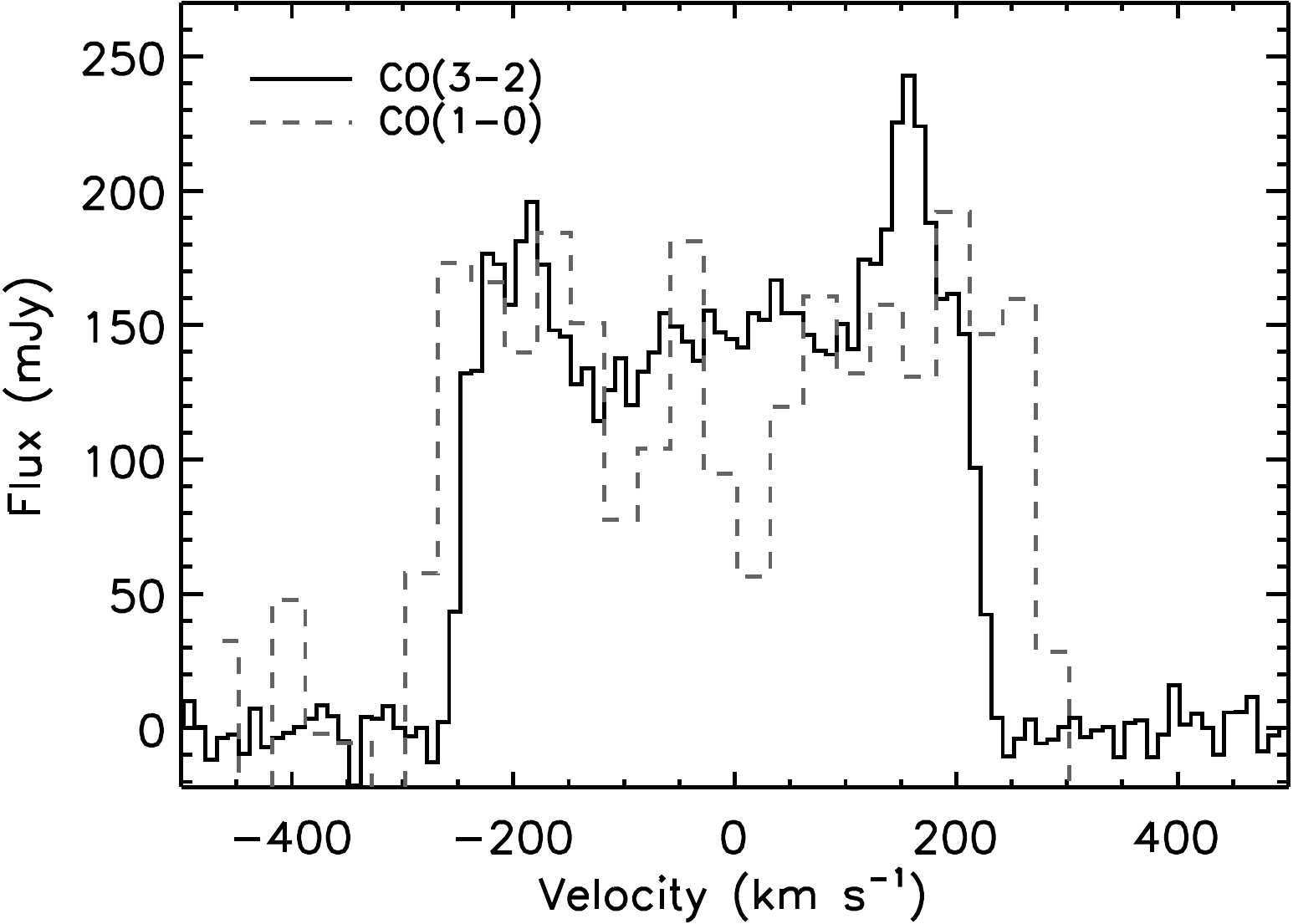}
\caption{Integrated $^{12}$CO(3-2) spectrum extracted from our observed data cube in an 8\arcsec$\times$4\arcsec\ (440 pc $\times$ 220 pc) region around the galaxy centre, covering all the detected emission. The spectrum shows the classic double-horn shape of a rotating disc. Also shown as a dashed grey histogram is the CARMA $^{12}$CO(1-0) integrated spectrum from \protect \cite{2013MNRAS.432.1796A}, that has a significantly broader velocity width than that of the CO(3-2) emission.}
\label{spectrumplot}
 \end{center}
 \end{figure}

Figure \ref{spectrumplot} shows the integrated $^{12}$CO(3--2) spectrum of NGC4429, exhibiting the classic double-horn shape of a rotating disc. The total flux is 75.45~$\pm$~0.09~$\pm$7.5 Jy \kms\ (where the second error is systematic and accounts for the $\approx$10\% flux calibration uncertainty of the ALMA data). 
By comparing this measurement with that of $^{12}$CO(1--0) \citep[][also shown as a grey dashed line in Figure \ref{spectrumplot}]{2013MNRAS.432.1796A}, we find a $^{12}$CO(3-2)/$^{12}$CO(1-0) intensity ratio of 1.06\,$\pm$\,0.15 (when using beam temperature units; see Fig \ref{spectrumplot}), within the range usually found for the discs of nearby spiral galaxies \citep[0.5--1.4, e.g.][]{2012ApJ...746..129T}. {The line emission we detect provides an estimate of the mass of the cold gas and its distribution. This may need to be taken when estimating the SMBH mass, as is discussed further in Section \ref{gas_mass_discuss}.}
The $^{12}$CO(1--0) line is substantially wider than the $^{12}$CO(3-2) line, suggesting they are distributed somewhat differently. This is discussed in detail in Section \ref{gasmorph}.

\begin{figure} \begin{center}
\includegraphics[width=0.48\textwidth,angle=0,clip,trim=0cm 0cm 0cm 0.0cm]{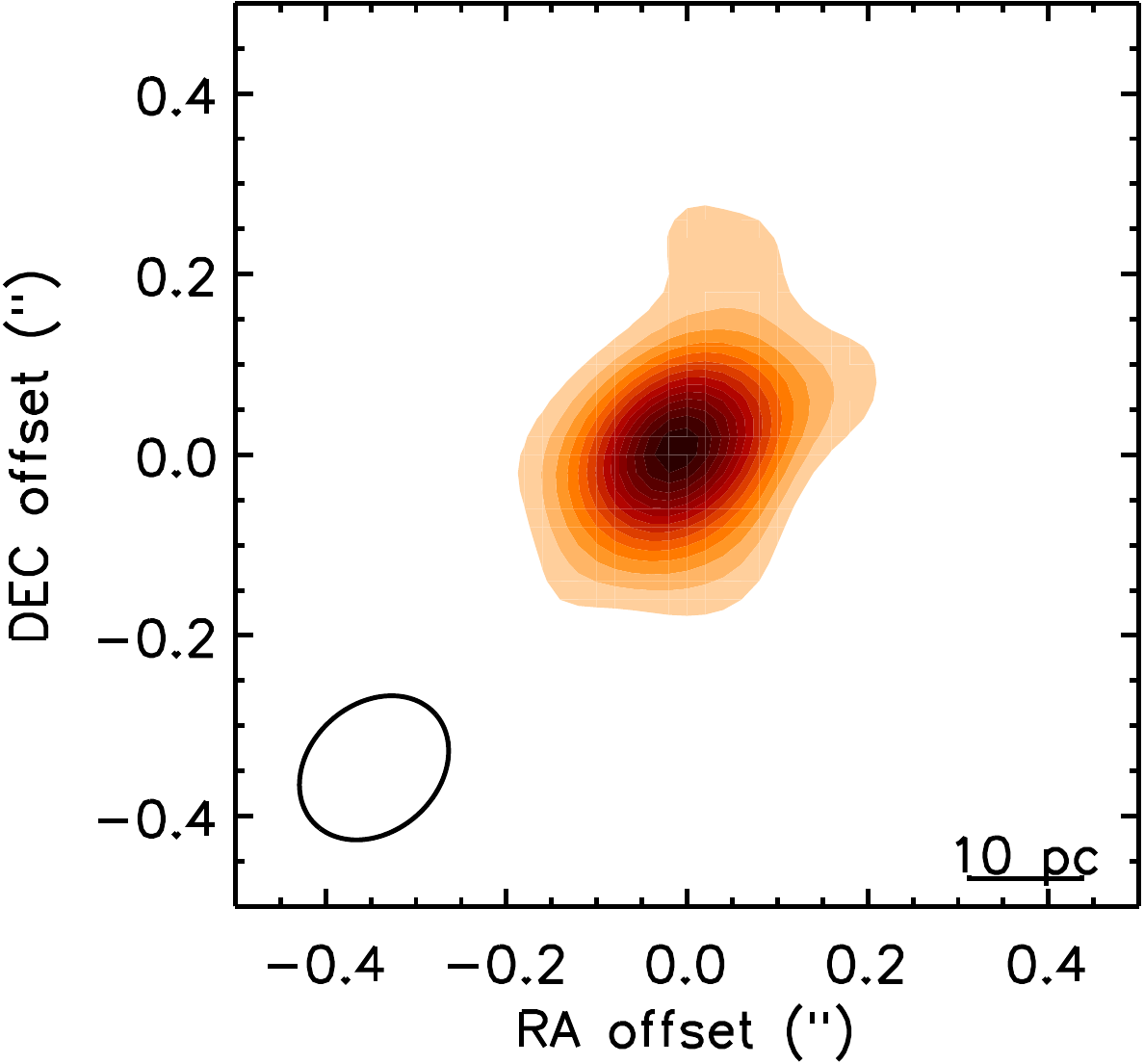}
\caption{Map of the 0.85 mm continuum emission in NGC4429. The contours plotted range from 87.5\,$\mu$Jy to 1.05\,mJy with steps of 35$\mu$Jy (2.5$\sigma$ to 30$\sigma$ in steps of 1$\sigma$). The synthesised beam is shown as an ellipse at the bottom left of the panel (0\farc18\,$\times$\,0\farc14 or 14\,$\times$\,11 pc$^2$). This emission is marginally resolved, and likely arises from dust around the nucleus of NGC4429.}
\label{contfig}
 \end{center}
 \end{figure}

 \subsection{Continuum emission}

As mentioned above, continuum emission was detected at the very centre of NGC4429 (Figure \ref{contfig}).
The peak of this continuum emission is detected at $>$30$\sigma$, and we find a total integrated intensity of 1.38\,$\pm$\,0.07\,$\pm$0.14 mJy (where again the second error is systematic and accounts for the ALMA flux calibration uncertainty). We constrain the position and size of this source using Gaussian fitting with the \texttt{CASA} task \texttt{IMFIT}. The emission is centred at an RA of 12$^{\rm h}$27$^{\rm m}$26\coordsec504\,$\pm$\,0\coordsec013 and a DEC of +11$^{\circ}$06$^{\prime}$27\farc57\,$\pm$\,0\farc01, where here the error bars include both the fitting errors and ALMA's astrometric accuracy. This position is consistent with the optical centre of NGC4429 (12$^{\rm h}$27$^{\rm m}$26\coordsec508\,$\pm$\,0\coordsec5, +11$^{\circ}$06$^{\prime}$27\farc76\,$\pm$\,0\farc5) as derived in the Sloan Digital Sky Survey (SDSS) data release 6 \citep{2008ApJS..175..297A}.

\texttt{IMFIT} shows that this source at the heart of NGC4429 is marginally spatially resolved. 
The deconvolved size (based on a gaussian source model) is estimated to be 112.2\,$\pm$\,16 by 63\,$\pm$\,16 milliarcseconds$^2$ ($\approx\,$9$\,\times\,$5\,pc$^2$), with a position angle of 143 $\pm$ 16 deg. We discuss the mechanism likely producing this emission in Section \ref{contdiscuss}.

\section{\uppercase{Method \& Results}}
 \label{method}
  \label{results}

Our goal in this paper is to estimate the mass of the SMBH at the centre of NGC4429. We are hindered, however, by the central hole present in the gas distribution (see Figure \ref{comomentfig}). This hole has a radius of $\approx$40\,pc, somewhat larger than the formal sphere of influence (SOI) of the expected SMBH ($R_{\rm SOI}\approx$ 15~pc given the source properties discussed above).
 However, as discussed in \cite{2014MNRAS.443..911D} and confirmed by our previous WISDOM works, the formal SOI criterion is not particularly meaningful for SMBH mass estimates using molecular gas kinematics (see also \citealt{2017MNRAS.466.1987Y}). When using a method based on cold gas, it is the central circular velocity profile (rather than the velocity dispersion profile) of the galaxy that is important.
 
We thus applied the figure of merit criterion ($\Gamma_{\rm FOM}$), as presented in Equation 4 of \cite{2014MNRAS.443..911D}, to our data. We used the black hole mass and channel size discussed above. 
 In NGC4429 the expected circular velocity at the edge of the central molecular hole due to the stellar mass distribution is 120~\kms, calculated by multiplying the luminous mass model of \cite{2013MNRAS.432.1894S} with the mass-to-light ratio of \cite{2013MNRAS.432.1709C}, and calculating the circular velocity as described in Appendix A of \cite{2002MNRAS.333..400C}. The errors in the circular velocities due to our mass model are assumed to be $\pm$10~\kms (see Section \ref{uncertainties}). Ellipse fitting to the dust disc in the F606W band \textit{HST} image yields an inclination $i$\,=\,66$^{\circ}$. 
When combined, these parameters yield a $\Gamma_{\rm FOM}$ of 3, suggesting we should be able to put constraints on the SMBH mass in this system. We note, however, that due to the central hole in the gas distribution, any SMBH mass estimate will be sensitive to errors in our stellar mass modelling (as discussed further below). 

In the rest of this Section, we therefore outline our method to estimate the SMBH mass and other physical parameters of NGC4429 from the observed molecular gas kinematics.

\subsection{Gas dynamical modelling}

In this paper we use the same forward modelling approach applied in the other papers of this series \citep{2017arXiv170305247O,2017arXiv170305248D} to estimate the black hole mass (and other physical parameters) of NGC4429. 

Briefly, we utilise the publicly available KINematic Molecular Simulation (\textsc{KinMS}\footnote{https://github.com/TimothyADavis/KinMS}) mm-wave observation simulation tool of \cite{2013MNRAS.429..534D}, coupled to the Markov Chain Monte Carlo (MCMC) code \textsc{KinMS\_mcmc} (Davis et al. in prep.). This tool allows input guesses for the true gas distribution and kinematics and, assuming the gas is in circular rotation, produces a simulated data cube that can be compared to the observed data cube (taking into account the observational effects of beam-smearing, spatial and velocity binning, disc thickness, gas velocity dispersion, etc.).
Full details of the fitting procedure are described in WISDOM paper II \citep{2017arXiv170305248D}, but we highlight the important aspects below.

\subsubsection{Gas distribution}

One of the inputs of the \textsc{KinMS} models is an arbitrarily parameterised function $\Sigma_{\rm gas}(r)$ that describes the radial gas surface brightness distribution, here assumed to be axisymmetric.
Because the molecular disc in NGC4429 is clumpy (due to the detection of individual molecular clouds) we do not aim to reproduce all the internal features of gas distribution. Instead, only the coarser features of the gas disc are fitted. 

\begin{figure} \begin{center}
\includegraphics[height=6.7cm,angle=0,clip,trim=0cm 0cm 0cm 0.0cm]{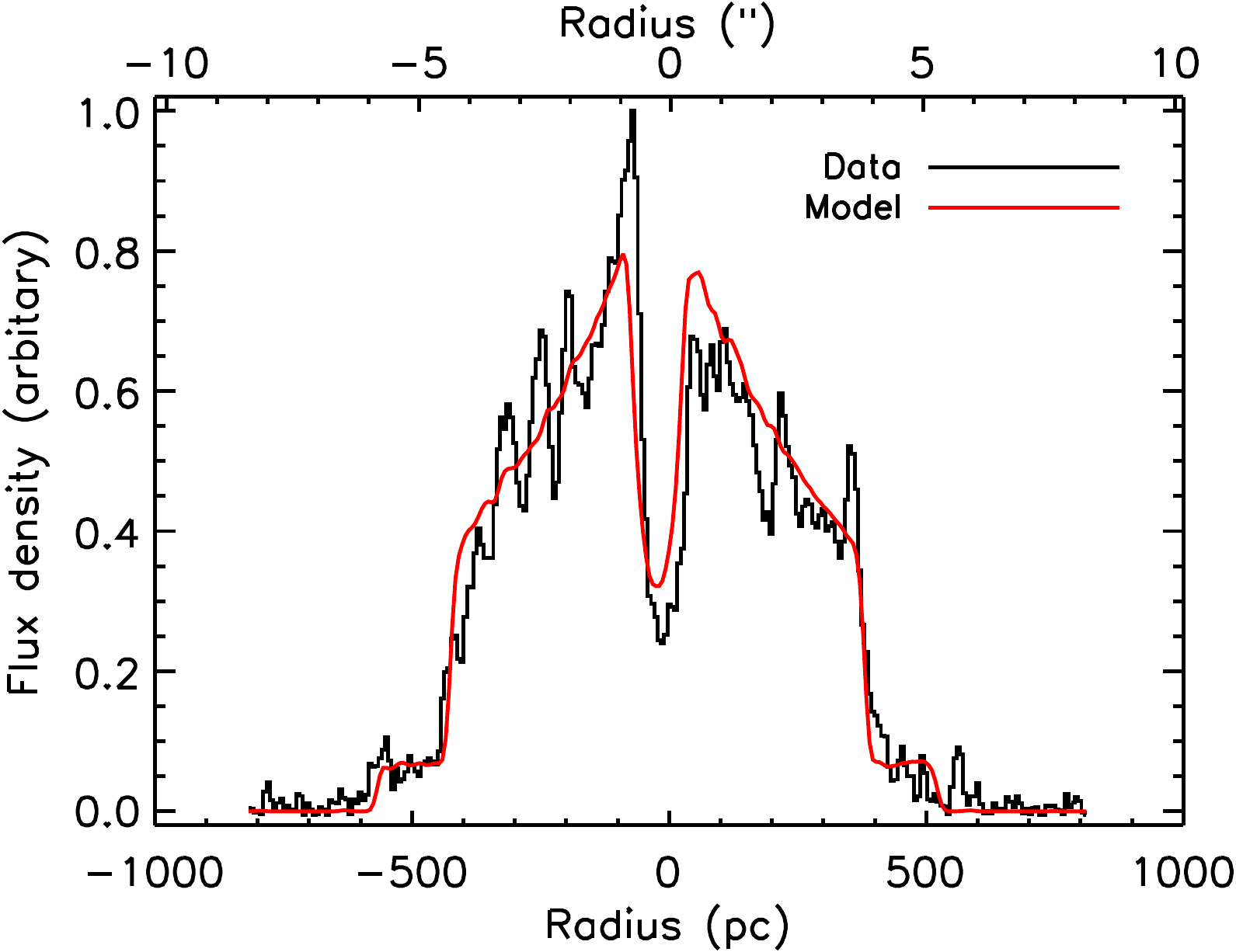}
\caption{A major axis cut through the NGC4429 integrated intensity map, showing the data in black and our best-fit \textsc{KinMS} model (with a surface brightness profile as defined in Eqn. \protect \ref{eq:surfbrightprof}) in red. Note that the gas surface density does not decrease smoothly to our detection limit, but instead appears to be truncated. }
\label{radialcut}
 \end{center}
 \end{figure}

As Figure \ref{comomentfig} shows, the gas disc in NGC4429 has a hole in the centre. In addition, our ALMA observations reveal that the gas surface density does not decrease smoothly to our detection limit, but instead appears to be truncated (see Figure \ref{radialcut}). We thus model the gas distribution using an exponential disc with inner and outer truncations. There are a few scattered molecular clouds that extend slightly further out and these we model with a constant low surface brightness plateau. This distribution has 5 free parameters: the exponential disc scale length ($r_{\rm scale}$), the inner and outer truncation radii ($r_{\rm inner}$ and $r_{\rm outer}$), the plateau surface brightness $\Sigma_{\rm plateau}$ and the outer edge radius $r_{\rm plateau}$. Overall the profile is parameterised as

\begin{equation}
\Sigma_{\rm gas}(r)\propto 
\begin{cases}
    \,0\,,              & \text{if } r < r_{\rm inner}\\
        \,\exp{\left(\frac{-r}{r_{\rm scale}}\right)} + \Sigma_{\rm plateau}\,,& \text{if }  r_{\rm inner} \leq r\leq r_{\rm outer}\\
           \, \Sigma_{\rm plateau}\,, & \text{if } r_{\rm outer} < r \leq r_{\rm plateau}\\
    \,0\,,              & \text{if } r > r_{\rm plateau}\\
\end{cases}.
\label{eq:surfbrightprof}
\end{equation}

\noindent The first truncation provides the inner hole in the gas distribution. The second provides the sharp edge in the outer parts of the CO(3-2) distribution (see Fig. \ref{radialcut}). The plateau allows us to model the few low-surface brightness clouds that extend beyond the sharp edge of the disc. We note that excluding the plateau would not change any of the derived model parameters. The best-fit profile, convolved to match our observations, is shown in red in Figure \ref{radialcut}.

Various other free parameters of the gas disc are also included in the model. These are the total flux, position angle ($\phi_{\rm kin}$), and inclination ($i$) of the gas disc, as well as its kinematic centre (in RA, Dec, and velocity). We find no evidence of a warp in this galaxy, so the inclination and position angle are each fit with a single value valid throughout the disc. Overall, our gas distribution model thus has 11 free parameters.

\begin{figure*} \begin{center}
\includegraphics[height=6.5cm,angle=0,clip,trim=0cm 0cm 0cm 0.0cm]{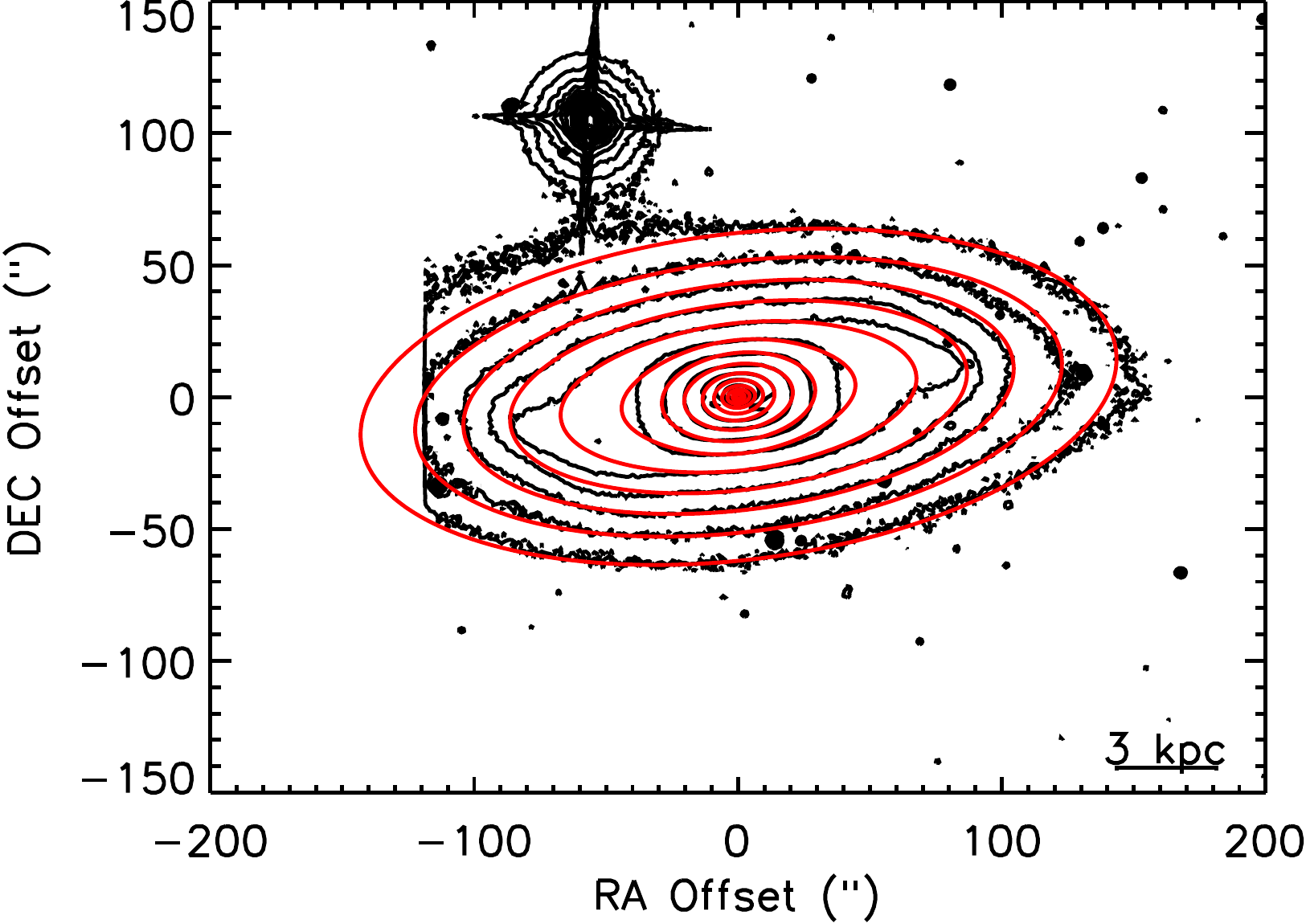}
\includegraphics[height=6.5cm,angle=0,clip,trim=0cm 0cm 0cm 0.0cm]{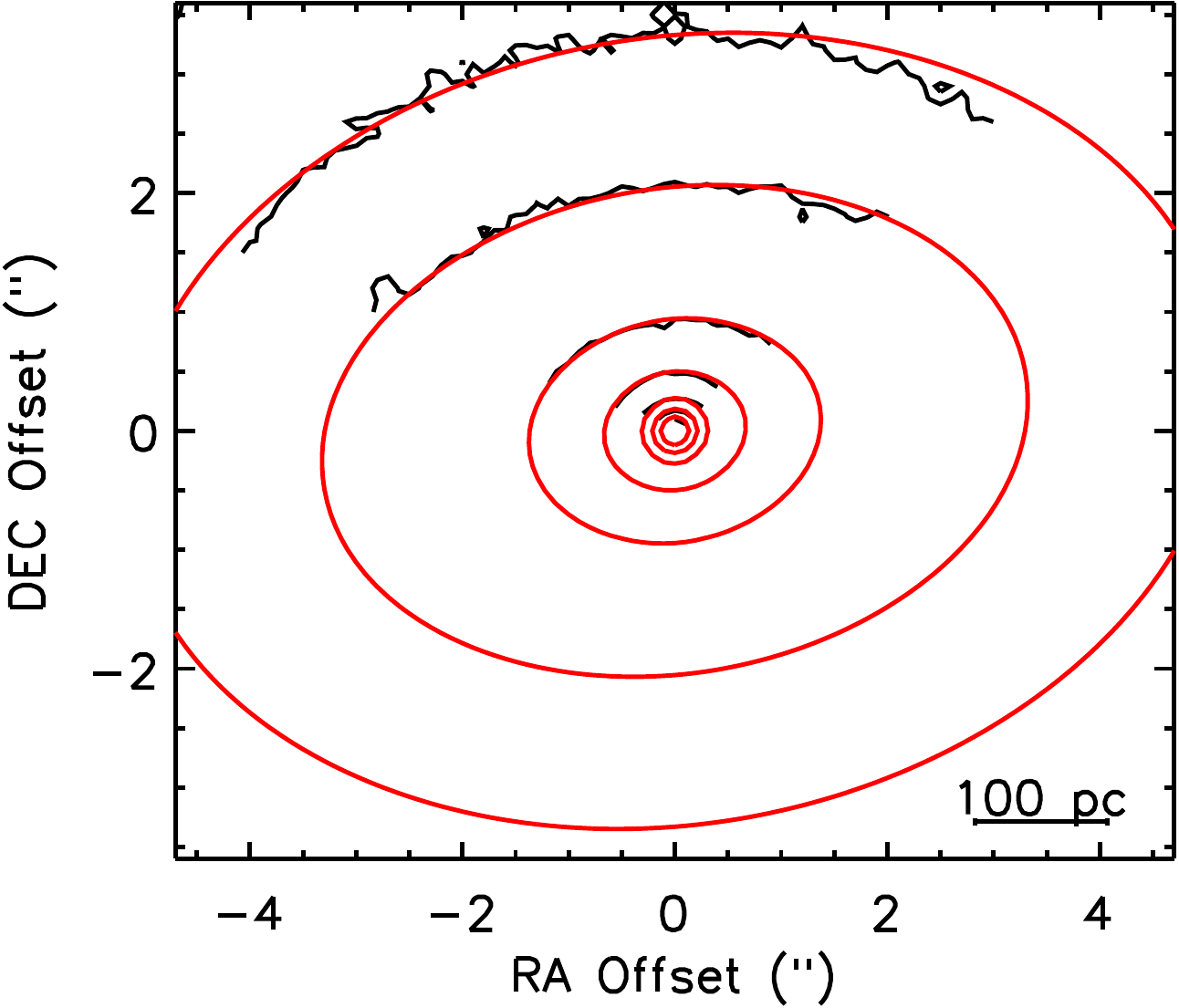}
\caption{Our MGE model of NGC4429 (red contours), overlaid on an \textit{SDSS} $r'$-band image on the left and an \textit{HST} Wide-Field Planetary Camera 2 (WFPC2) F606W image on the right (both shown with black contours). The small central dust disc has been masked in the fitting procedure. Note that we do not attempt to fit the boxy bulge or spiral features, but rather the smooth underlying disc structure.}
\label{mge_galplot}
 \end{center}
 \end{figure*}

\subsubsection{Stellar distribution}
\label{stellerdist}

\begin{figure} \begin{center}
\includegraphics[height=6.cm,angle=0,clip,trim=0cm 0cm 0cm 0.0cm]{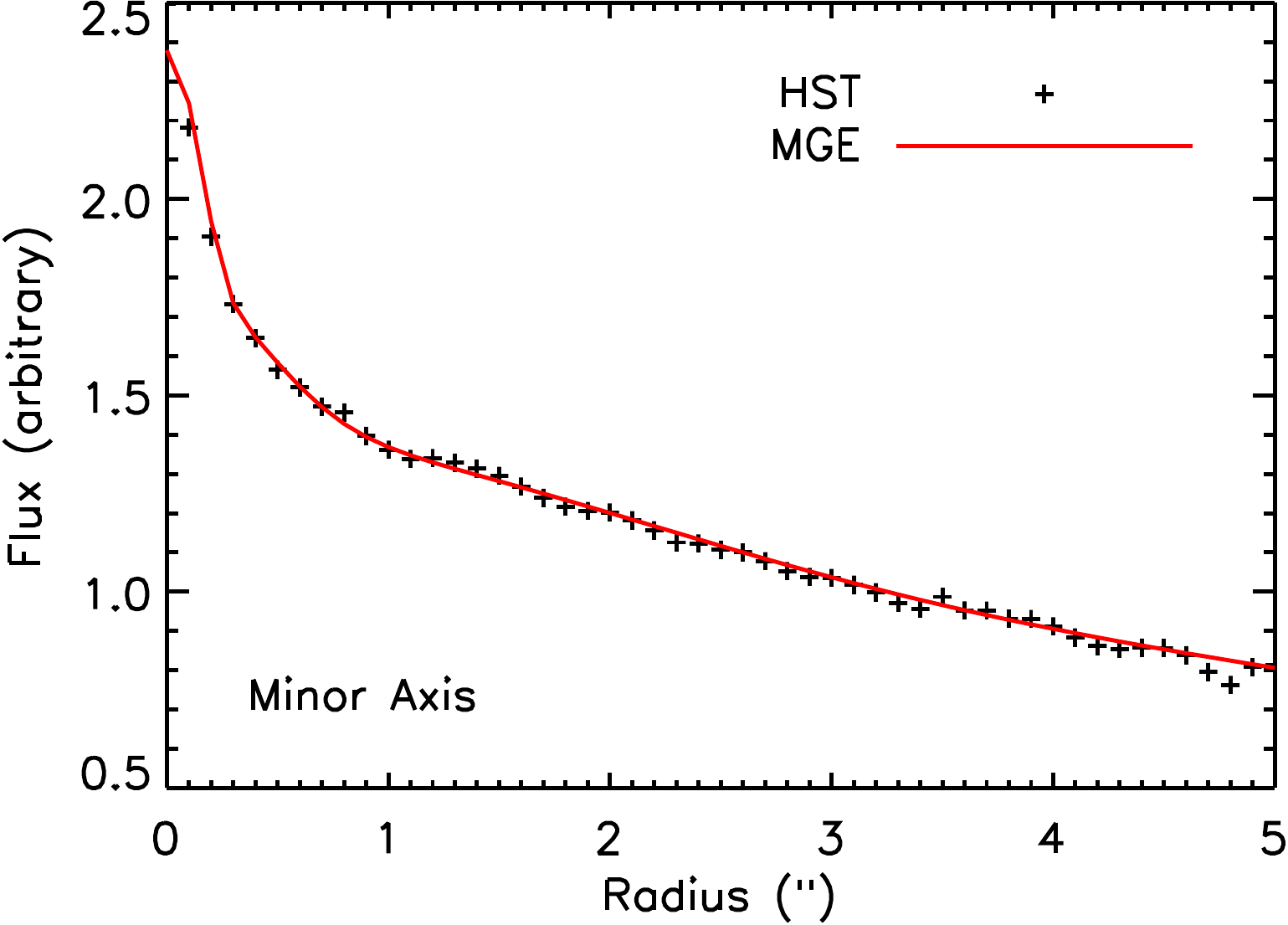}
\caption{Radial surface brightness distribution of the \textit{HST} Wide-Field Planetary Camera 2 (WFPC2) F606W image of NGC4429 along its minor axis (black crosses). In red is our MGE model, demonstrating that we achieve a good fit to the inner light distribution.}
\label{mge_galplot_minor}
 \end{center}
 \end{figure}

As in our other works, to account for the contribution of visible matter and calculate the true mass of any dark object present at the centre of NGC4429, we construct a luminous mass model. 
We parameterise the luminous matter distribution using a multi-Gaussian expansion (MGE; \citealt{Emsellem:1994p723}) model of the stellar light distribution, constructed using the \textsc{MGE\_FIT\_SECTORS} package of \cite{2002MNRAS.333..400C}.
Our best-fit MGE model is tabulated in Table \ref{mgetable} and shown visually in Figure \ref{mge_galplot}.
 This was constructed from an \textit{HST} Wide-Field Planetary Camera~2 (WFPC2) F606W image, in combination with an $r'$-band image from SDSS. We used the F606W image as our reference for flux calibration, assuming the absolute magnitude of the Sun in this band to be 4.67 \citep{2012PASP..124..606M}. 
As in our previous works we mask the region where the central dust disc obscures the stellar light profile, leaving only the unobscured minor axis to constrain our fits. This masking makes it hard to judge the quality of our fits from the images in Figure \ref{mge_galplot}, and thus we also show in Figure \ref{mge_galplot_minor} a cut along the minor axis of the \textit{HST} image of NGC4429, overlaid with the MGE profile. This demonstrates that we achieve an excellent fit to the inner light distribution.
 
We note that our MGE models assume that the object is axisymmetric, so we do not attempt to fit the boxy/spiral features at intermediate radii in NGC4429, but rather only the smooth underlying disc structure of this fast-rotating ETG. As the CO is confined to the innermost kiloparsec of this object, we do not expect this to affect our results.

\begin{table}
\caption{MGE parameterisation of the galaxy F606W light profile.}
\begin{center}
\begin{tabular*}{0.4\textwidth}{@{\extracolsep{\fill}}r r r}
\hline
$\log_{10}$ I$_j$ & $\log_{10}$  $\sigma_j$ & $q_j$ \\
$L_{\odot,F606W}$ pc$^{-2}$ & (") & \\
(1) & (2) & (3)\\
\hline
$^{*}$5.402 &  -1.144 & 0.622 \\
3.896 &  -0.290 & 0.666 \\
3.516 &   0.385 & 0.579 \\
3.176 &   0.717 & 0.700 \\
2.795 &   1.133 & 0.620 \\
2.284 &   1.691 & 0.400 \\
1.533 &   2.026 & 0.451 \\
 \hline
\end{tabular*}
\parbox[t]{0.45\textwidth}{ \textit{Notes:} For each Gaussian component, Column 1 lists its F606W surface brightness, Column 2 its standard deviation (width) and Column 3 its axis ratio. The central unresolved Gaussian, indicated with a star, is removed to minimise the effect of the AGN on our kinematic fitting.}
\end{center}
\label{mgetable}
\end{table}%

This model of the stellar light can be de-projected analytically, given an inclination that we fix here to be the same as that of the molecular disc (as fitted by \textsc{KinMS\_mcmc}). 
When combined with a stellar mass-to-light ratio ($M$/$L$; another free parameter in our modelling), the light model then directly predicts the circular velocity of the gas caused by the luminous matter. 
We note that NGC4429 contains an AGN, that likely contributes significantly to the unresolved point source at the galaxy centre. We subtract this point source by removing the innermost (spatially unresolved) Gaussian from our MGE model (as listed in Table \ref{mgetable}). 
We note that including this point source does not significantly alter our results (as discussed in Section \ref{mlgradsec}).

Here we find (as in \citealt{2016arXiv160903559D}) that a $M$/$L$ gradient is required to fit the observed gas kinematics. After testing various functional forms (e.g. linear, second order power law, broken power law), we found that assuming a $M$/$L$ (referred to here as $\Psi$ for brevity) that is piecewise linear as a function of radius provides the best fit to the data. We thus parameterise $\Psi(r)$ as

\begin{equation}
\Psi(r)= 
\begin{cases}
   \, m_1r+\Psi_{\rm cent}\,, & \text{if } r \leq r_{\rm break}\\
   \\
    \,m_2r+c_2\,,& \text{if }  r_{\rm break} \leq r < r_{\rm outer}\\
\end{cases},
\label{eq:mlgrad}
\end{equation}
\noindent where
\begin{eqnarray}
m_1 = \frac{\Psi_{\rm break} - \Psi_{\rm cent}}{r_{\rm break}}\,,\\
m_2 = \frac{\Psi_{\rm outer}-\Psi_{\rm break}}{r_{\rm outer}-r_{\rm break}}\,,\end{eqnarray}\\
\noindent and
\begin{eqnarray}
c_2=\Psi_{\rm break}-m_2r_{\rm break}\,.
\label{intercept2}
\end{eqnarray}

$\Psi_{\rm cent}$ is the the $M$/$L$ in the galaxy centre, $\Psi_{\rm break}$ and $r_{\rm break}$ are the $M$/$L$ and radius of the break in the M/L profile, respectively,  and $\Psi_{\rm outer}$ is the $M$/$L$ at the edge of our detected gas disc ($r_{\rm outer}$=\,5\arcsec\ or 400\,pc). As we define the $M$/$L$ to be a continuous function, the intercept of the outer component is entirely determined by $\Psi_{\rm cent}$, $m_2$ and $r_{\rm break}$, as shown in Equation \ref{intercept2}. 
In what follows, however, we fix the inner component to be flat ($m_1$ = 0, $\Psi_{\rm break} = \Psi_{\rm cent}$) as our data always drove the fit towards this solution. 
Using this form we thus add three free parameters to our fitting. 
We discuss this choice of parameterisation in detail in Section \ref{mlgradsec}, but note here that all reasonable forms tested that did not include a centrally peaked mass concentration (that leads to a Keplerian-like rise in the circular velocity curve) could not reproduce our data without the addition of a SMBH (see Section \ref{uncertainties}). 

To include this $M$/$L$ gradient in our model we follow \cite{2016arXiv160903559D} by scaling by $\sqrt{\,\Psi(r)}$ the circular velocity curve calculated from our MGE model assuming an $M$/$L$ of unity.  
The derived $\Psi(r)$ is valid in the F606W band, as defined in the \textit{HST} WFPC2 system, although we abbreviate this to $M$/$L_V$ here.

In this work we always assume that the gas is in circular motion, and hence that the gas rotation velocity varies only radially. 
We do, however, also include a parameter for the internal velocity dispersion of the gas, that is assumed to be spatially constant. 
Note that our treatment of the velocity dispersion is not self-consistent dynamically (i.e. we do not solve Jeans' equations, but simply add a Gaussian scatter to our velocities). 
As the gas velocity dispersion is much smaller than its circular velocity at all radii in molecular gas discs (and indeed here), this is very unlikely to affect our results.

As already mentioned in Section \ref{lineemission}, the gas velocity dispersion in this source appears to be significantly smaller than our channel width of 10 \kms. We conduct an independent analysis with our 2 \kms\ channel$^{-1}$ cube in Section \ref{veldisp1} to estimate the true value of the gas velocity dispersion. We henceforth fix this parameter to that value ($\approx$2.2 \kms) in our SMBH mass fitting. We tested that allowing this parameter to vary during SMBH fitting does not alter any of the other best fit parameters.
The effects of other potential deviations from circular motion are discussed in Section \ref{uncertainties}.

\subsection{Bayesian analysis}
\label{fitting}

As mentioned above, we use a Bayesian analysis technique to identify the best set of model parameters. This allows us to obtain samples drawn from the posterior distribution of the fifteen model parameters (including the SMBH mass; see Table \ref{fittable}). We utilise a MCMC method with Gibbs sampling and adaptive stepping to explore the parameter space. 
A full description of this MCMC code as well as details of cross checks with the well tested MCMC code \textsc{emcee} \citep{2013PASP..125..306F} will be presented in Davis et al., in preparation.

As our data are approximately Nyquist sampled spatially, neighbouring spaxels are strongly correlated by the synthesised beam. To deal with this issue we calculate the full covariance matrix analytically, and include it when estimating the likelihood. We use this covariance matrix with a standard logarithmic likelihood function based on the $\chi^2$ distribution. Full details of this procedure are described in Section 4.2.1 of \cite{2017arXiv170305248D}. The disadvantage of this method is that the numerical inversion of the covariance matrix becomes unstable for large images. We thus limit ourselves to fitting only 64\,$\times$\,64 pixel$^2$ areas when including the full covariance calculation.

\begin{table*}
\caption{Best-fit model parameters and statistical uncertainties.}
\begin{center}
\begin{tabular*}{0.8\textwidth}{@{\extracolsep{\fill}}l r c r r r r}
\hline
Parameter & \multicolumn{3}{c}{Search range} & Best fit & Error (68\% conf.) & Error (99\% conf.)\\
(1) &  \multicolumn{3}{c}{(2)} & (3) & (4) & (5)\\
\hline
Black hole:&&&&&& \\\hline
log$_{10}$ SMBH mass (M$_{\odot}$) &   4.8& $\rightarrow$ &  9.8 &     8.17 &$\pm$0.01 & $\pm$0.03\\
\\
Stars: &&&&&&\ \\\hline
Inner stellar $M$/$L$ ($\Psi_{\rm cent}$; M$_{\odot}$/L$_{\odot,V}$) &   1.0& $\rightarrow$ & 10.0 &     6.59 & $\pm$0.05 & -  0.13, +  0.12\\
Inner $M$/$L$ break radius ($r_{\rm break}$; ") &   1.0& $\rightarrow$ &  4.0 &     1.43 & -  0.05, +  0.06 & -  0.16, +  0.12\\
Outer Stellar $M$/$L$ ($\Psi_{\rm outer}$; M$_{\odot}$/L$_{\odot,V}$)$^*$ &   1.0& $\rightarrow$ & 10.0 &     8.25 & -  0.03, +  0.02 & $\pm$0.06\\
\\
Molecular gas disc: &&&&&&\ \\\hline
Kinematic position angle ($\phi_{\rm kin}$; $^\circ$) &   0.0& $\rightarrow$ &359.0 &    93.20 & $\pm$0.03 &$\pm$0.09\\
Inclination ($i$; $^\circ$) &  66.5& $\rightarrow$ & 89.0 &    66.80 & $\pm$0.04 & -  0.15, +  0.14\\
Disc scale length ($r_{\rm scale}$; \arcsec) &   0.0& $\rightarrow$ & 10.0 &     4.67 & $\pm$0.04 & $\pm$0.12\\
Central hole radius ($r_{\rm inner}$; \arcsec) &   0.0& $\rightarrow$ &  1.0 &     0.61 & $\pm$0.01 & -  0.03, +  0.02\\
{Outer luminosity cut radius ($r_{\rm outer}$; \arcsec)}$^*$ &   4.0& $\rightarrow$ &  8.0 &     5.09 & $\pm$0.03 & $\pm$0.08\\
{Plateau outer radius ($r_{\rm plateau}$; \arcsec)}$^*$ &   4.0& $\rightarrow$ & 10.0 &     6.86 & $\pm$0.03 & $\pm$0.09\\
{Plateau surface brightness} ($\Sigma_{\rm plateau}$)$^*$ &   0.0& $\rightarrow$ &  0.2 &     0.08 & $\pm$0.01 & $\pm$0.02\\
\\
Nuisance parameters: &&&&&&\ \\\hline
Luminosity scaling &   1.0& $\rightarrow$ &150.0 &    57.91 & -  0.45, +  0.42 & -  1.47, +  1.29\\
Centre X offset (\arcsec) &  -1.0& $\rightarrow$ &  1.0 &    -0.17 & $\pm$0.01 & $\pm$0.01\\
Centre Y offset (\arcsec) &  -1.0& $\rightarrow$ &  1.0 &     0.04 & $\pm$0.01 &$\pm$0.01\\
Centre velocity offset (\kms) & -20.0& $\rightarrow$ & 20.0 &    -5.73 & -  0.06, +  0.05 & -  0.22, +  0.19\\ \hline
\end{tabular*}
\parbox[t]{0.8\textwidth}{ \textit{Notes:} Column 1 lists the fitted model parameters, while Column 2 lists the prior for each. The prior is assumed to be uniform in linear space (or in logarithmic space for the SMBH mass only). The posterior distribution of each parameter is quantified in the third to fifth columns (see also Fig. 3). The parameters indicated with a star ($^*$) in Column 1 were constrained in the fit to the full data cube only, and hence their uncertainties are likely to be underestimated. The X, Y and velocity offset nuisance parameters are defined relative to the ALMA data phase centre position (12$^{\rm h}$27$^{\rm m}$26\coordsec508, +11$^{\circ}$06$^{\prime}$27\farc76, V=1104 \kms).}
\end{center}
\label{fittable}
\end{table*}

\subsubsection{Fitting process}

As the molecular gas disc in NGC4429 is large compared to the synthesised beam, we adopt a two-step fitting process. We initially fit a 160\,$\times$\,160 pixel$^2$ area that roughly covers the entire molecular gas disc, using a simple $\chi^2$-based likelihood function without taking into account the covariance between pixels. This allows us to obtain estimates of the parameters that are constrained by the outer disc ($\Psi_{\rm outer}$, $r_{\rm outer}$, $\Sigma_{\rm plateau}$, $r_{\rm plateau}$). Ignoring the covariance between pixels does not affect the best-fit values, but it does mean their uncertainties will be somewhat underestimated. However, because constraining these parameters is not the goal of this work, and the SMBH mass does not depend on them, we do not expect this simplification to affect our conclusions. Once these four parameters have been determined we fix their values and run the fitting a second time, including the covariance matrix (now with only 11 free parameters), but only in the central 64\,$\times$\,64 pixel$^2$ (3\farc2\,$\times$\,3\farc2) area of the galaxy disc.

We set reasonable flat priors (an assumption of maximal ignorance) on all the free parameters during both of these fitting procedures to ensure our kinematic fitting process converges. These are listed in Table \ref{fittable}. The kinematic centre of the galaxy was constrained to lie within two beam-widths of the optical galaxy centre position. The systemic velocity was allowed to vary by $\pm$20 \kms\ from that found by optical analyses. The disc scale length was constrained to be less than 10\arcsec. The $M$/$L$ parameters were constrained such that the $M$/$L$ could vary between 0.1 and 10.0 \msun/L$_{\odot,V}$ within the inner 5\arcsec of the galaxy. The prior on the SMBH mass was flat in log-space, with the mass allowed to vary between log$_{10}(\frac{M_{\rm BH}}{\mathrm{M}_{\odot}})$~=~4.8 and 9.8. Good fits were always found well within these ranges. The inclination of the gas disc was allowed to vary over the full physical range allowed by the MGE model. As is often the case, the best-fit inclination lies near the edge of this allowed range (suggesting the flattest Gaussians in our MGE are indeed tracing a disc-like structure at this inclination). 
{Both the inclination and position angle priors were flat in angle, rather than in the sine/cosine quantities that actually enter the analysis. We re-ran the fitting process with flat priors in sin($i$) and cos($\phi_{\rm kin}$), respectively, and confirmed that this does not affect the best-fit values obtained.}

\begin{landscape}
\begin{figure} \begin{center}
\includegraphics[width=23cm,angle=0,clip,trim=0cm 0cm 0cm 0.0cm]{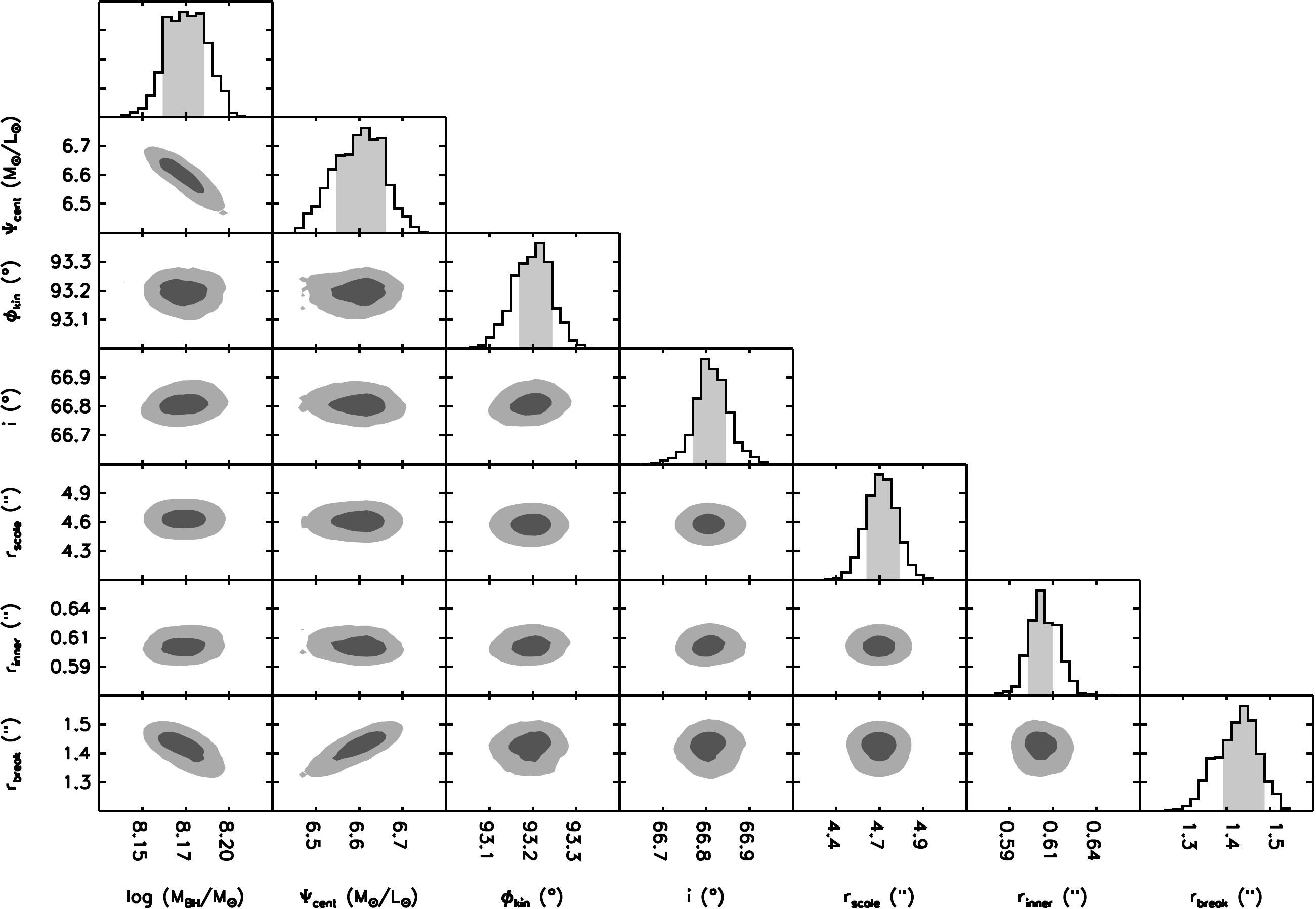}
\caption{
Visualisation of the multidimensional parameter space explored by our fit to the observed data from the central 3\farc2\,$\times$\,3\farc2 of NGC4429. In the top panel of each column a one-dimensional histogram shows the marginalised posterior distribution of that given parameter, with the 68\% (1$\sigma$) confidence interval shaded in pale grey. In the panels below, the greyscale regions show the two-dimensional marginalisations of those fitted parameters. Regions of parameter space within the 99\% confidence interval are coloured in pale grey, while regions within the 68\% confidence interval are coloured in dark grey. See Table \ref{fittable} for a quantitative description of the likelihoods of all fitting parameters.}
\label{triangleplot}
 \end{center}
 \end{figure}
\end{landscape}

Both fits were run until convergence, and then the best chains were run for 100,000 steps (with a 10\% burn-in) to produce our final posterior probability distributions. For each model parameter these probability surfaces were then marginalised over to produce a best-fit value (the median of the marginalised posterior samples) and associated 68\% and 99\% confidence levels (CLs). Figure \ref{triangleplot} shows the one- and two-dimensional marginalisations of the physical parameters included in the fit to the galaxy central regions (SMBH mass, M$_{\rm BH}$; $M$/$L$ central value, $\Psi_{\rm cent}$; break radius, $r_{\rm break}$; molecular gas exponential disc scale, $r_{\rm scale}$; disc truncation radius, $r_{\rm inner}$; disc inclination, $i$; and disc position angle, $\phi_{\rm kin}$). Quantitative descriptions of the likelihoods of all parameters are presented in Table \ref{fittable}.  

Most of our parameters are independent of each other, but we do find some degeneracies. As expected, the SMBH mass is degenerate with $\Psi_{\rm cent}$ (as is always the case in any SMBH mass fit). The inner parameters of the $M$/$L$ profile ($\Psi(r)$) are also somewhat degenerate, as $\Psi_{\rm cent}$ correlates positively with $r_{\rm break}$. Despite these degeneracies all three parameters remain well constrained by the data, and the additional scatter introduced is included when marginalising to obtain our final uncertainties.

 \begin{figure*} \begin{center}
\includegraphics[height=5.5cm,angle=0,clip,trim=0cm 0cm 0cm 0.0cm]{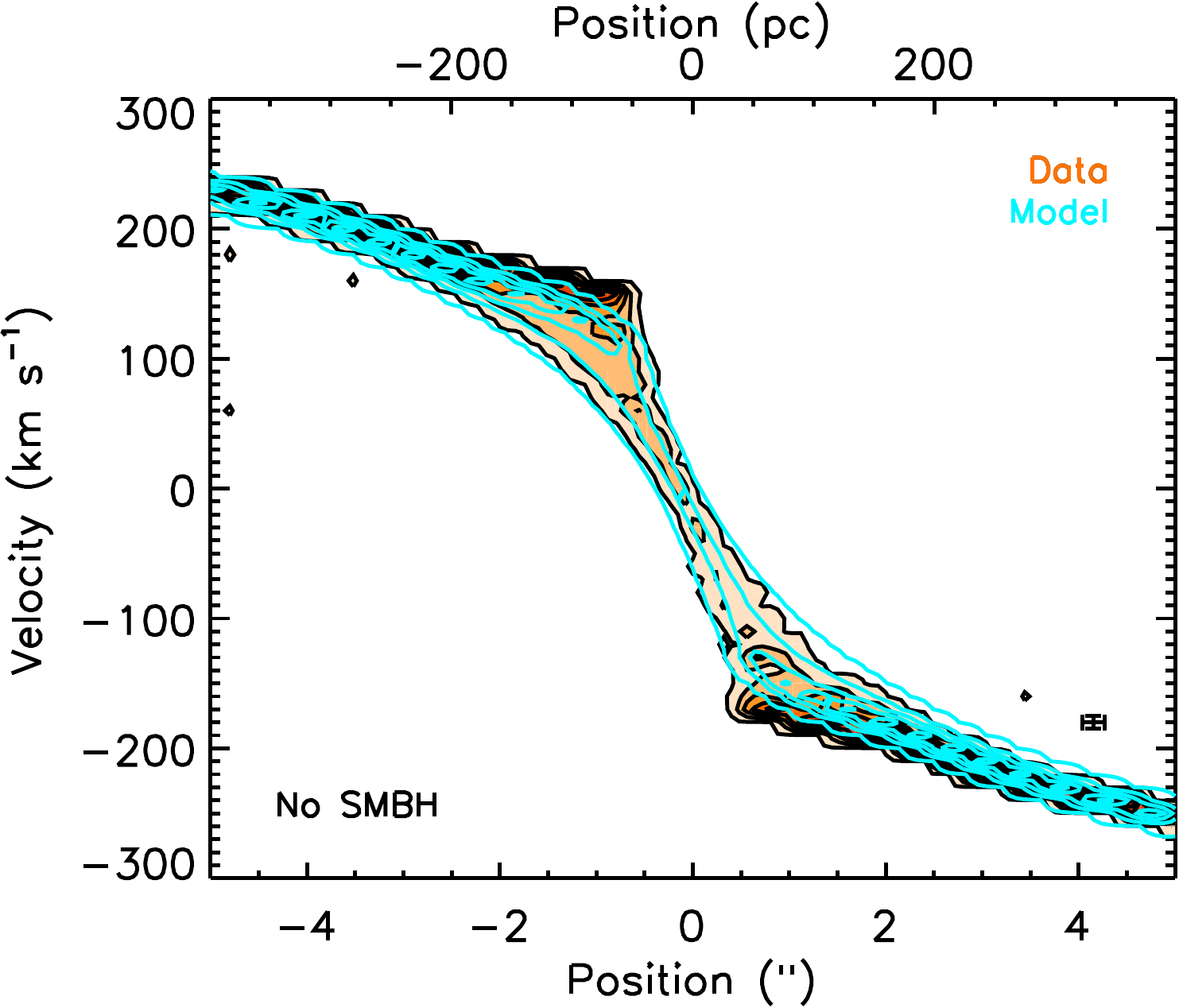}\includegraphics[height=5.5cm,angle=0,clip,trim=2.5cm 0cm 0cm 0.0cm]{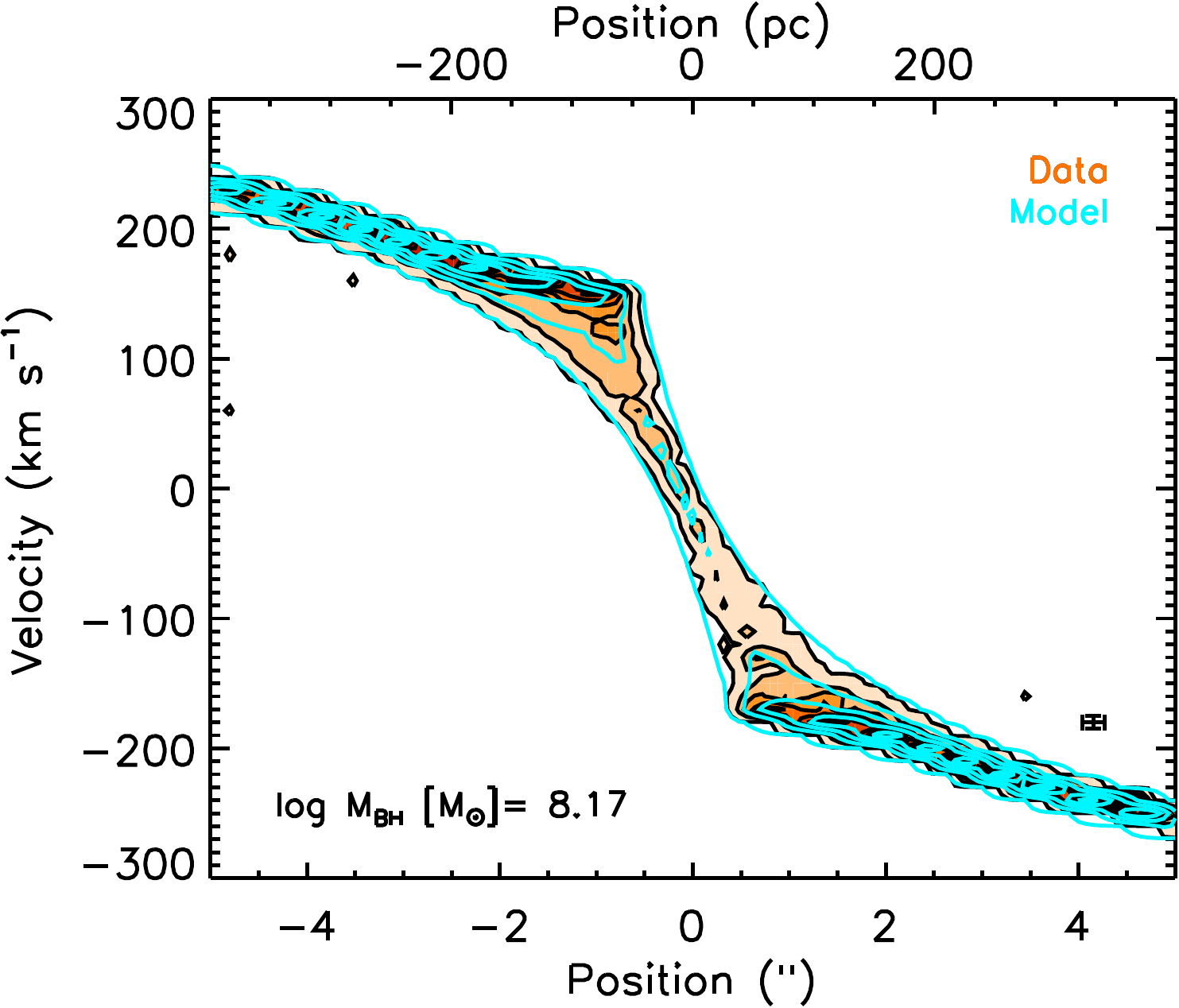}\includegraphics[height=5.5cm,angle=0,clip,trim=2.5cm 0cm 0cm 0.0cm]{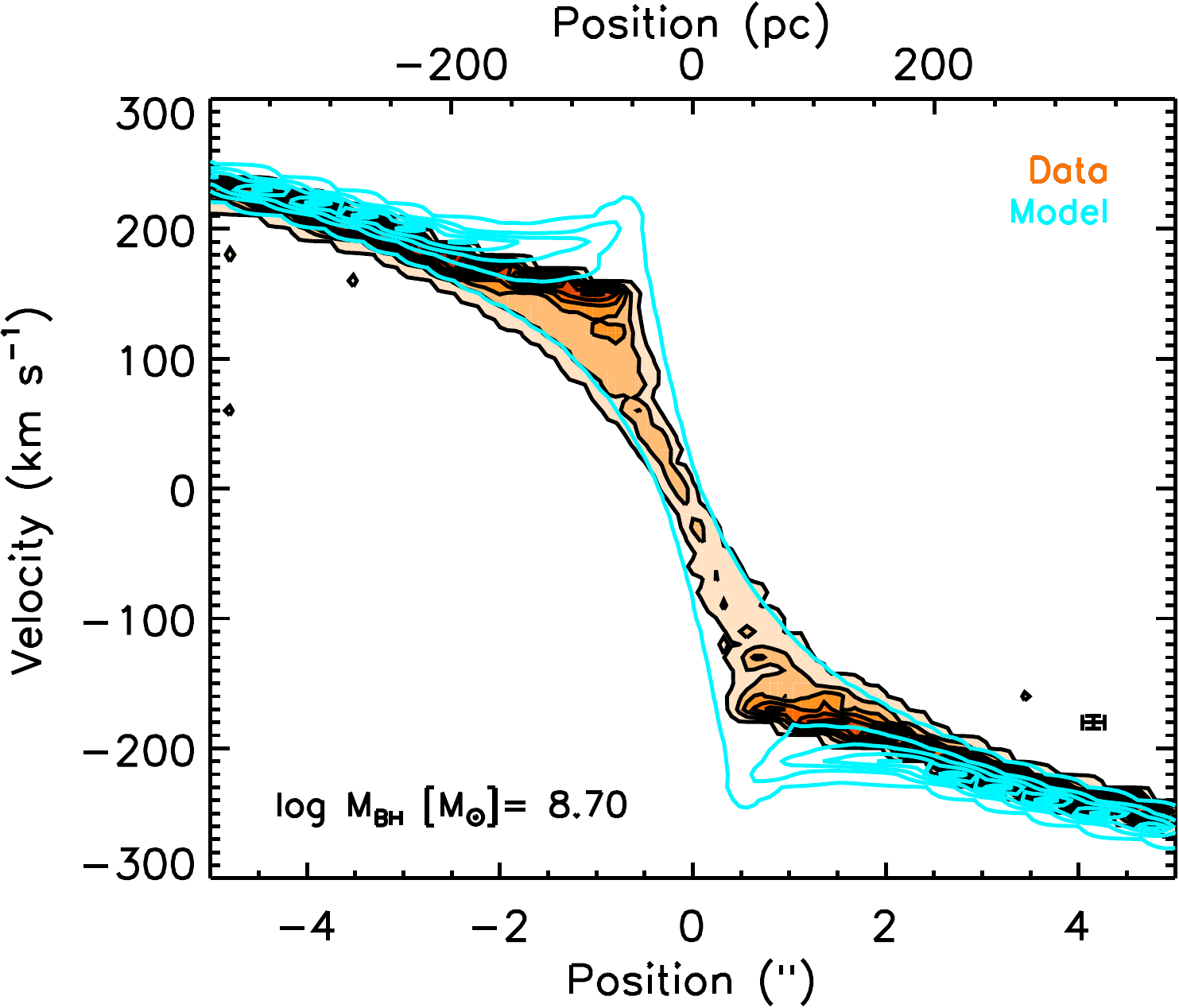}
\caption{Major-axis PVD of the $^{12}$CO(3-2) emission in NGC4429 (orange scale and black contours), as in Figure \ref{pvdplot}, overlaid with model PVDs extracted in an identical fashion from models that only differ by the central SMBH mass (blue contours). The left panel has no SMBH, the centre panel shows our best-fit SMBH mass, and the right panel has an overly massive SMBH. The legend of each panel indicates the exact SMBH mass used. The error bars in the bottom right of the plots show our beam and channel width. A model with no SMBH is clearly not a good fit to the data in the central parts.  }
\label{threepvdplot}
 \end{center}
 \end{figure*}

\begin{landscape}
\begin{figure}
 \begin{center}
\includegraphics[width=24cm,angle=0,clip,trim=0cm 0cm 0cm 0.0cm]{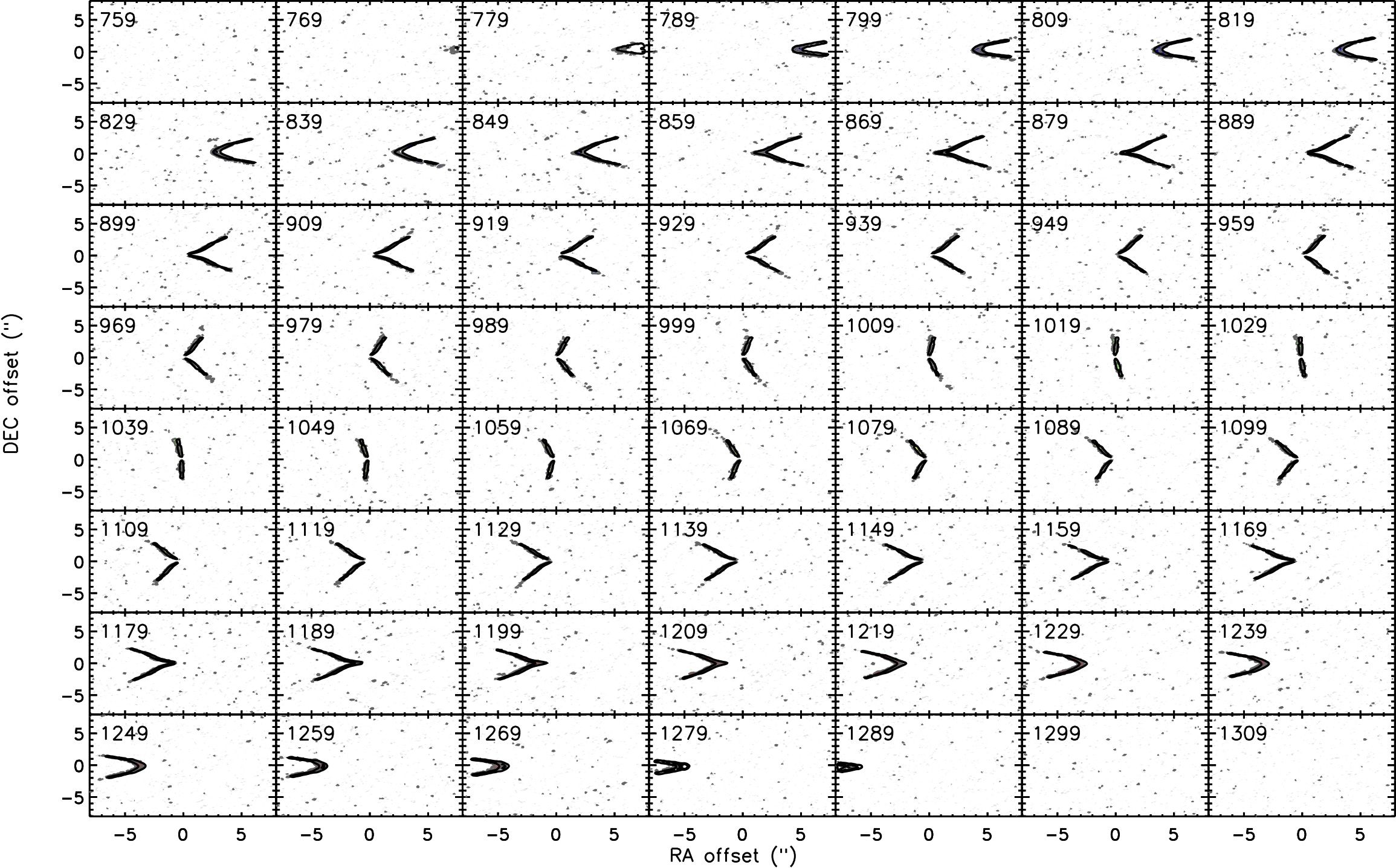}
\caption{Channel maps of our ALMA $^{12}$CO(3-2) data in the velocity range where emission is detected. The coloured regions with grey contours show the areas detected with more than 2.5$\sigma$ significance. Overplotted in black are the same contour levels from our best-fit model. Our model agrees well with the observed data in every velocity channel.}
\label{fitpar_compare_chanmap}
\label{channelmaps}
 \end{center}
\end{figure}
 \end{landscape}

We clearly detect the presence of a massive dark object in the centre of NGC4429, with a mass of (1.5$\pm$0.1) $\times$10$^8$ \msun\ (at the 99\% CL).
Figure \ref{threepvdplot} shows the observed PVD overlaid with the best-fit model, and with models with no SMBH and an overly large SMBH. A model with a fine-tuned SMBH mass clearly reproduces the kinematics of the molecular gas better than the alternatives.
In Figure \ref{fitpar_compare_chanmap} we show our best-fit model (in black) overlaid on the channel maps of our ALMA observations. The model is an excellent fit to the gas kinematics throughout the molecular gas disc.

 The best-fit $V$-band mass-to-light ratio is 6.6$_{-0.10}^{+0.09}$ \msun/L$_{\odot,V}$ in the central parts of the galaxy, rising to 8.25$\pm0.06$ \msun/L$_{\odot,V}$ at 400~pc.  Note that the uncertainties quoted here include random errors only, but we discuss systematic sources of error in Section \ref{uncertainties}.
 These $M$/$L$s are somewhat different from the $r'$-band value of 6.1$\pm$0.4 (within $\approx$1\,R$_{\rm e}$ or 3.2 kpc) derived via Jeans Anisotropic Models (JAM; \citealt{2008MNRAS.390...71C}) of the SAURON stellar kinematics by \cite{2013MNRAS.432.1709C}. Note however that the JAM model in that work has low quality (qual=0), due to the dust features preventing an accurate MGE fit from the SDSS image alone, especially in the region of the dust disc (see \citealt{2013MNRAS.432.1894S}). We performed a new mass-follows-light JAM fit on the SAURON data using the MGE given in Table 1 and find $M$/$L_{r}$=8.8, in better agreement with the gas dynamical values presented here.

\subsection{Main uncertainties}
\label{uncertainties}

Sources of uncertainty in molecular gas SMBH mass measurements have been considered in depth in previous papers of this series, and we refer readers to these works for full details. In this Section we briefly touch upon the main uncertainties that could affect the specific results derived in this work.
 
\subsubsection{Distance uncertanties}

All SMBH mass estimates are systematically affected by the distance they assume to their target galaxy (with $M_{\rm BH} \propto D$). For NGC4429 we use a surface brightness fluctuation distance from \cite{Tonry:2001ei}, that has a formal uncertainty of $\approx$10\%. Thus the distance-related systematic uncertainty on the SMBH mass is of a similar order as our random errors. We include this systematic uncertainty estimate as a second error term on our SMBH mass in what follows.

\subsubsection{Mass model uncertainties}
\label{mlgradsec}
We use an MGE model of the stellar light distribution to estimate the contribution of luminous matter to the observed gas kinematics.
This model can be contaminated by dust and the presence of an AGN, and the conversion between light and mass is also uncertain.  This could, in principle, lead to an under/overestimate of the stellar luminosity/mass in some parts of our object, and thus bias the derived parameters.

A potential contaminant is the optically obscuring dust that is clearly present in the centre of NGC4429 (see Figure \ref{gal_overview}).
To minimise the effect of this dust on our MGE model we carefully mask the affected regions to avoid any contamination. This has been shown to work well to recover the intrinsic light distribution of such systems \citep{2002MNRAS.333..400C}. In addition, we are helped here by the fact that the nucleus of the galaxy appears relatively extinction free (within $\approx$0\farc5). Dust is present in most of the molecular disc, however, where the $M$/$L$ is primarily constrained. 

A changing $M$/$L$ with radius might be expected within the star-forming molecular disc. \cite{2016arXiv160903559D} studied this in detail in a sample of ETGs (including NGC4429). Although, in general, they found such gradients to be negligible, NGC4429 
was identified as having a significant gradient. We thus explicitly included and fit for the radial variation of the $M$/$L$ here. 

\begin{figure} \begin{center}
\includegraphics[width=0.48\textwidth,angle=0,clip,trim=0cm 0cm 0cm 0.0cm]{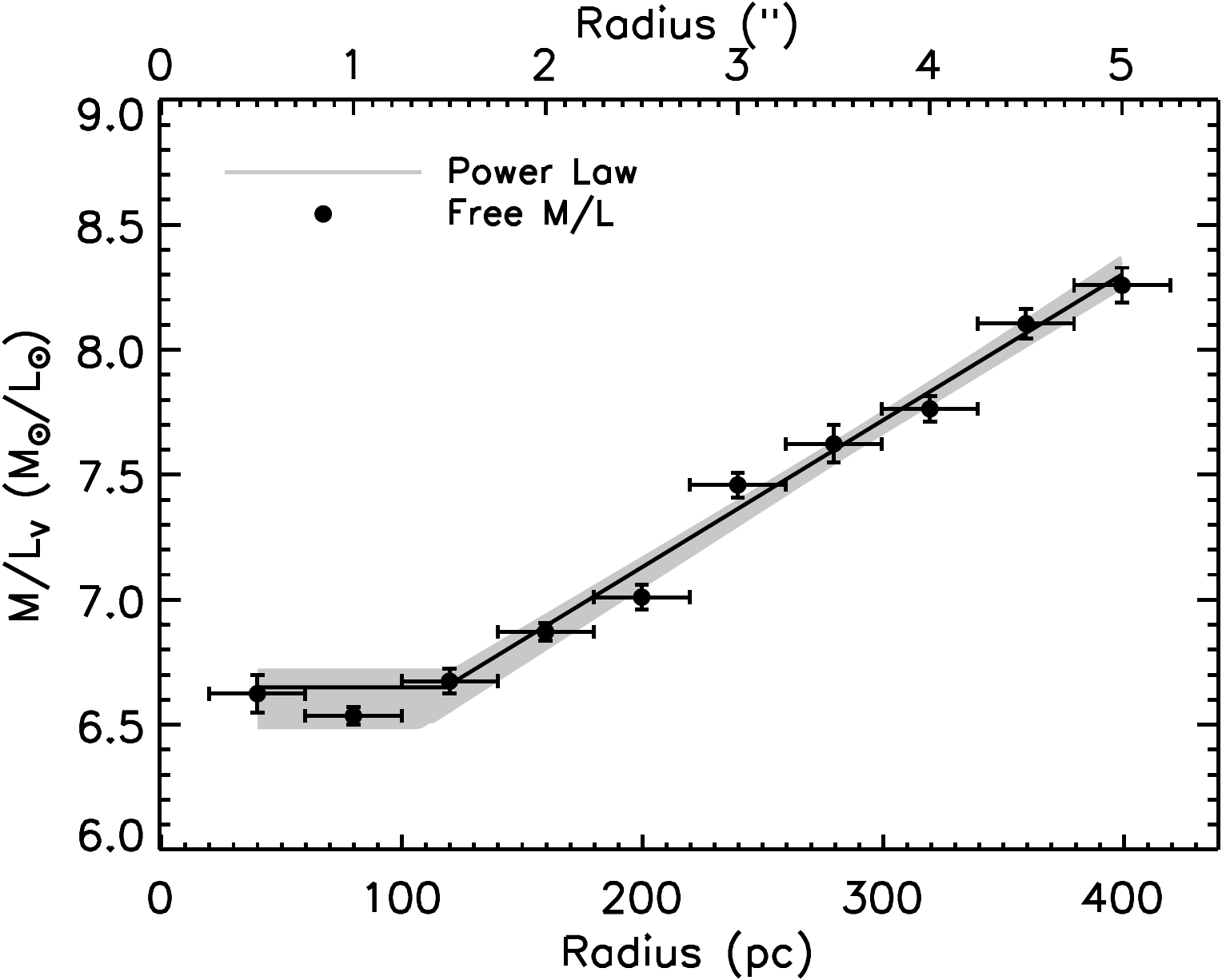}
\caption{Mass-to-light ratio radial profile in NGC4429. The black line is our best-fit piecewise linear function, while 200 realisations drawn from our posterior probability distribution are shown in grey, revealing the allowed range of variation around this best fit. The black points with error bars are a second independent test, where the $M$/$L$ was allowed to vary independently in 0\farc5\,width bins. Both methods agree within the errors, suggesting a piecewise linear function is a reasonable match to the data.}
\label{ml_grad}
 \end{center}
 \end{figure}

Figure \ref{ml_grad} shows the best-fit $M$/$L$ radial profile in our object (black line), along with 200 realisations drawn from our posterior probability distribution (grey curves), showing the allowed range of variation around this best fit. To independently verify that a simple piecewise linear model is a good representation of the data we repeated our fitting procedure, allowing the $M$/$L$ to vary independently in 10 radial bins (following the formalism introduced by \citealt{2016arXiv160903559D}). The black points with error bars in Figure \ref{ml_grad} show these estimates, that agree well with our simple piecewise linear approximation. This gives us confidence that such a model, although not unique, at least well reproduces the features seen in our data.

This approach also has the advantage that we are able to subsume the effect of any simple error in the mass profile of the galaxy into our fitted $M$/$L$ gradient. For instance, the dust appears fairly uniform in our \textit{HST} imaging, and thus the $M$/$L$ gradient we allow for should allow us to marginalise over its potential effect. Our previous papers highlighted the possibility of errors in the MGE mass profile itself, and these can also be subsumed as long as they can be reasonably fitted by our model.
 
{As mentioned above, given the central hole in the gas distribution, any SMBH mass estimate will be sensitive to errors in our stellar mass modelling and determination of the $M$/$L$ gradient. Indeed, our best-fit model predicts a total stellar mass of 3$\times10^8$ \msun\ within $r_{\rm inner}$ (the radius at which the molecular hole begins), and thus our SMBH makes up only 1/3 of the total mass in that region.}
It is thus interesting to consider what sort of $M$/$L$ variation would remove the need for an SMBH entirely.  Matching the velocity of the gas around the edge of the central molecular gas hole is possible, but it would require $\Psi_{\rm cent} \approx$11.6 $M_{\odot}$/$L_{\odot, V}$. 
However, simply boosting the central $M$/$L$ does not enable a good fit to the rest of the data cube to be found. The only way to match the observations is to include a rising central $M$/$L$ (that leads to a Keplerian-like rise in the circular velocity curve). Although we cannot rule out this possibility (as with all other SMBH mass estimates), the simplest way to create such a rise is the presence of a massive dark object such as an SMBH.

A further possible contaminant is the AGN. NGC4429 does have some central radio emission and a broad H$\alpha$ line that suggest ongoing low-level accretion. 
This AGN may contribute to the F606W waveband flux at the very centre, as a nuclear point source is seen in the image. 
We remove this point source from our MGE model in an attempt to minimise the effect of the AGN on our mass determination. 
Re-running the fitting procedures described above, but including the central unresolved component of our MGE model, yields a decrease of the SMBH mass of 0.2 dex. We include this additional uncertainty by adding it to our downward systematic error term in what follows. The true contribution of the AGN to the F606W flux is likely to be intermediate between the two cases tested, and thus we conclude that the presence of the AGN is unlikely to significantly bias our SMBH mass measurement.

\subsubsection{Fitting area}

As mentioned above (Section \ref{lineemission}), no gas is present at the very centre of NGC4429 (around the SMBH itself). Molecular gas is detected only at larger radii, where the potential of the SMBH is still important but it is harder to disentangle its effect from that of the stellar potential. We thus depend on our mass model being accurate enough to derive SMBH constraints.

When fitting the gas kinematics over the whole data cube, many resolution elements are available to help us constrain the $M$/$L$ and its gradient, but very few are available (at small radii) to constrain the SMBH mass. This can lead us to underestimate the uncertainty on the SMBH mass. To explore this effect, we repeated our fitting procedure using only data within 1\arcsec\ of the galaxy centre. We recovered the same best-fit SMBH mass and $M$/$L$, with increased uncertainties as expected (by a factor of $\approx$1.5 in this case). This suggests we are not biasing our best-fit SMBH mass by fitting a larger area of the molecular gas disc.

\subsubsection{Non-circular motions}

Significant non-circular motions (e.g. inflow, outflow, streaming) could affect our analysis, because we have assumed that the gas in NGC4429 is in purely circular motion. 
We consider this a reasonable assumption, however, because although this object clearly harbours a large-scale bar, this bar is unlikely to have a strong effect at small radii, deep within the bulge (where the molecular gas disc lies).  
 We can also test for the presence of such motions by looking at the residuals between our data and the best-fit model. No such strong residual is found in NGC4429, suggesting that there is no significant non-circular motion within its very dynamically cold disc.

\subsubsection{Gas velocity dispersion}
\label{veldisp1}
The molecular gas velocity dispersion provides an additional source of uncertainty \citep[see][]{2016ApJ...823...51B}. In the models presented here, we allow for a single value of the velocity dispersion throughout the disc. 
However, as mentioned in Section \ref{stellerdist}, the velocity dispersion is not well constrained by our data cube with a 10~\kms\ channel width.

To investigate what the true gas velocity dispersion is, we re-reduced our ALMA data with a channel width of 2~\kms. 
At this velocity resolution, we can constrain $\sigma_{\rm gas}$ if it is larger than 0.8 \kms. We repeated the kinematic fitting described above on this higher velocity resolution data cube, allowing $\sigma_{\rm gas}$ to vary. These fits were able to constrain the velocity dispersion, yielding a value of 2.2$^{+0.68}_{-0.65}$ \kms (at the 99\% CL). We therefore fixed the gas velocity dispersion to this value in the kinematic fitting presented above (Section \ref{fitting}). Arbitrarily increasing the assumed gas velocity dispersion by an order of magnitude has a negligible effect on our best-fit SMBH mass, and thus we consider our results robust to any remaining uncertainty in this parameter.
We nevertheless discuss the surprisingly low velocity dispersion of NGC4429 further in Section \ref{veldispdiscuss}.

\subsubsection{Gas mass and dark matter}
\label{gas_mass_discuss}
In addition to luminous matter, the mass of dark matter and any ISM material present also contributes to the total dynamical mass of a galaxy, and thus could bias our best-fit parameters.

Dark matter, while important at larger galactocentric distances, is negligible at the radii we are probing here. \cite{2013MNRAS.432.1862C} found that NGC4429 has a negligible dark matter fraction within one effective radius ($R_{\rm e}$), and we are working well within this (at $<R_{\rm e}$/8).

However, the ISM of NGC4429 does provide an additional source of dynamical mass. Because this object is in the Virgo cluster, it is very \hi\ poor \citep[$\ltsimeq10^7$ \msun;][]{2012MNRAS.422.1835S}. Thus the molecular mass dominates any ISM contribution. In ETGs the molecular reservoirs are usually dynamically unimportant due to their low gas fractions (see e.g. \citealt{2016arXiv160903559D}).  
Here, the total H$_2$ mass over the entire molecular gas disc is only $\approx10^8$\,\msun\ (assuming a standard Galactic conversion factor), similar to the SMBH mass but distributed over a much larger area, and very much less than the estimated stellar mass within the same region ($\approx$4.5$\times$10$^{9}$ \msun; estimated from our best-fit circular velocity curve at the edge of the $^{12}$CO(3-2) emitting disk, $R$=5\arcsec). 
In addition, due to the central hole in the gas distribution, the H$_2$ mass around the SMBH (within our inner resolution elements) is very low.

{To estimate the maximum possible mass contribution from molecular gas in the central regions where it is undetected (within $r_{\rm inner}$) we multiply the surface brightness sensitivity achieved in our observations (87.6 \msun\ pc$^{-2}$- assuming a Galactic $X_{\rm CO}$ conversion factor and a CO(3-2)/CO(1-0) ratio of unity) by the area of the observed hole. The $3\sigma$ upper limit to the mass present in this region is then 1.9$\times10^6$ \msun. Any change in the  CO(3-2)/CO(1-0) ratio or $X_{\rm CO}$ factor would thus have to be extremely significant (increases of a factor $>50$) before undetected gas in the circum-nuclear regions of the galaxy could contribute significantly to the dark object mass we measure.}


We ran a test to estimate the impact that including the gas mass in our modelling might have. \textsc{KinMS} allows the user to input the total gas mass and thus have the potential of the gas distribution included in the calculation of the circular velocity. In this case (even when giving the largest possible mass to the molecular component with a super-solar conversion factor), the best-fit parameters from Table \ref{fittable} are unchanged. We thus consider our results robust against this uncertainty.

\section{\uppercase{Discussion}}
 \label{discuss}

  \subsection{SMBH mass}

In earlier sections we estimated a mass of  (1.5$\pm0.1^{+0.15}_{-0.35}$) $\times$10$^8$~\msun\ for the SMBH in the centre of NGC4429 (where the last two values reflect the random and systematic uncertainties, respectively).
This is fully consistent with the previous upper limit of 1.8~$\times$~10$^8$~\msun\ set by \cite{2009ApJ...692..856B}.

\begin{figure*} \begin{center}
\includegraphics[width=0.78\textwidth,angle=0,clip,trim=0cm 0cm 0cm 0.0cm]{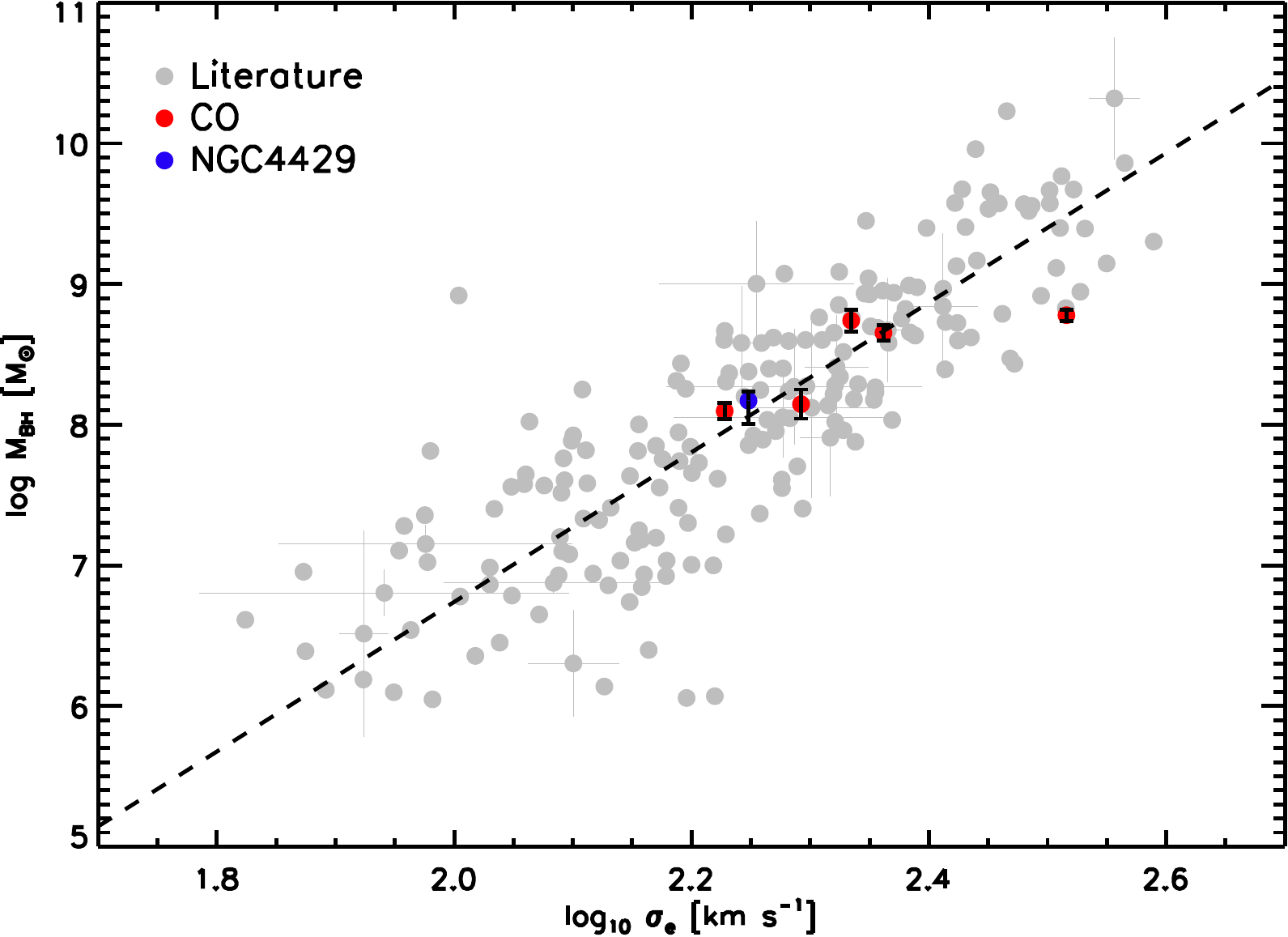}
\caption{$M_{\rm BH}$ -- $\sigma_{\rm e}$ relation from the compilation of \protect \cite{2016ApJ...831..134V} (grey points and dashed line). We show the SMBH mass measured for NGC4429 in this paper as a blue point, and highlight measurements from other works also using the molecular gas technique with red points. Our measurement of the SMBH mass of NGC4429 is consistent with the best-fit relation. }
\label{msigmaplot}
 \end{center}
 \end{figure*}

In Figure \ref{msigmaplot}, we show the location of our object on the $M_{\rm BH}$ -- effective stellar velocity dispersion ($\sigma_{\rm e}$) relation, using the compiled SMBH masses from \cite{2016ApJ...831..134V}.  NGC4429 is shown in blue and all of the other recent SMBH mass measurements using molecular gas are shown in red (\citealt{2013Natur.494..328D,2015ApJ...806...39O,2016ApJ...822L..28B,2017arXiv170305247O,2017arXiv170305248D}). NGC4429 is fully consistent with lying on the best-fit $M_{\rm BH}$ -- $\sigma_{\rm e}$ relation of \cite{2016ApJ...831..134V}. 
This suggests the presence of the bar in this system has not increased the average accretion rate onto the SMBH greatly, since it has not grown significantly beyond the $M_{\rm BH}$ -- $\sigma_{\rm e}$ expectation.

\subsection{Gas morphology and star formation}
\label{gasmorph}

\subsubsection{Radial extent}

The $^{12}$CO(3-2) emitting gas in NGC4429 is well fit by an exponential disc (with an exponential scale-length of 400\,$\pm$\,10\,pc), that is truncated at an inner radius of 48\,$\pm$\,3\,pc and an outer radius of 406\,$\pm$\,10\,pc. There are some scattered clouds that extend to a radius of $\approx$548\,$\pm$\,10 pc. The whole structure is likely to be in equilibrium, as the dynamical time at the outer edge of the disc is only 10 Myr.

\begin{figure} \begin{center}
\includegraphics[width=0.48\textwidth,angle=0,clip,trim=0cm 0cm 0cm 0.0cm]{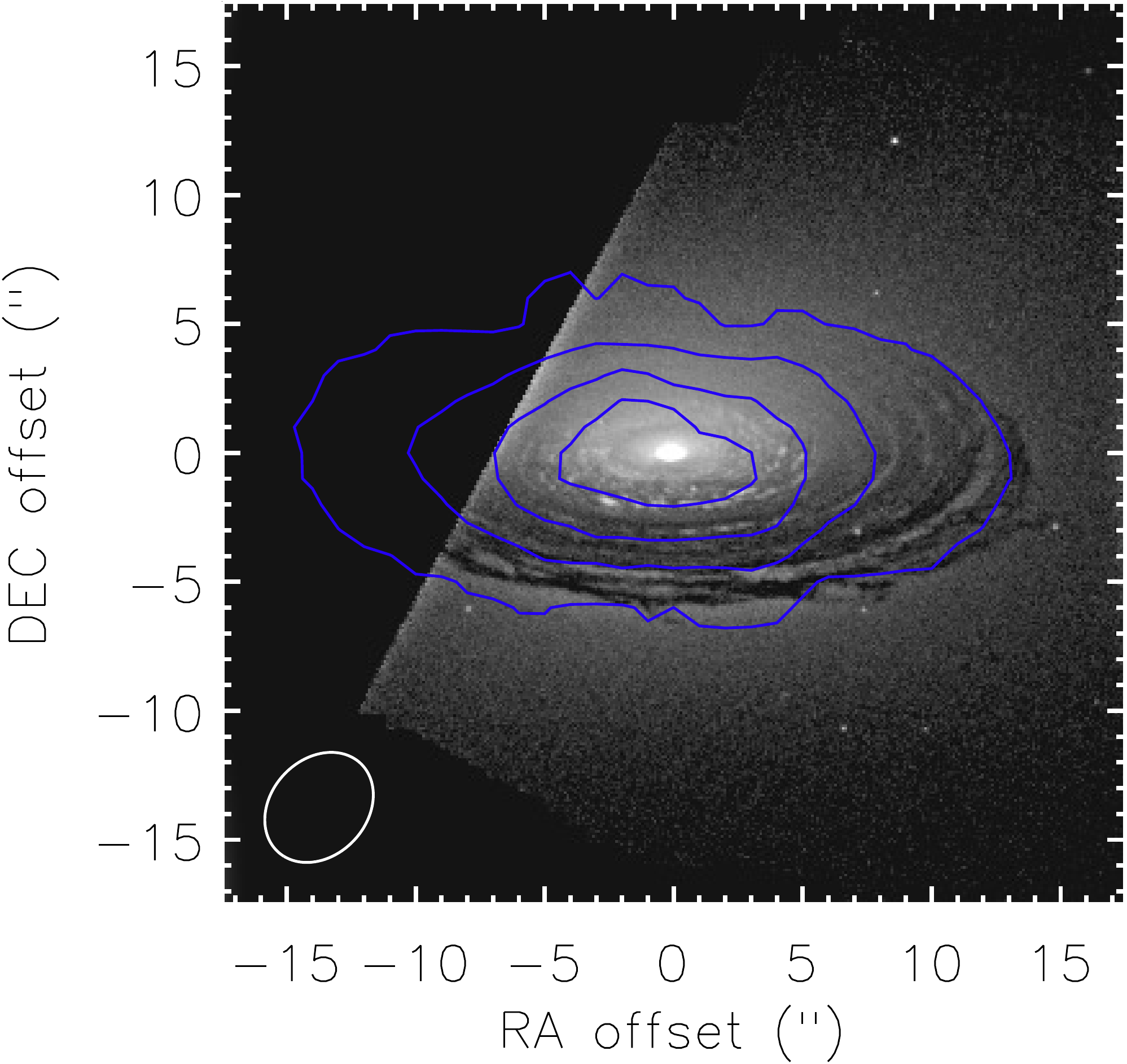}
\caption{CARMA $^{12}$CO(1-0) integrated intensity contours (blue), overlaid on the same unsharp-masked \textit{HST} WFPC2 F606W image as in Figure~\protect\ref{gal_overview}. Note that the $^{12}$CO(1-0) emitting gas extends to the edge of the dust disc, with a larger extent than the $^{12}$CO(3-2) emitting gas shown in Figure~\ref{gal_overview}.}
\label{carma}
 \end{center}
 \end{figure}

In Figure~\ref{carma} we show the integrated intensity map of $^{12}$CO(1-0), from CARMA observations of NGC4429 by \cite{2013MNRAS.432.1796A}, overplotted on the same \textit{HST} unsharp-masked image as in Figure~\ref{gal_overview}. Despite the vastly different resolutions, the CARMA $^{12}$CO(1-0) map is clearly inconsistent with the $^{12}$CO(3-2) map. This visual impression is confirmed by creating a \textsc{KinMS} model using the best-fit $^{12}$CO(3-2) surface brightness profile, but convolved with the CARMA synthesised beam. This does not provide a good fit to the $^{12}$CO(1-0) data cube.
The $^{12}$CO(1-0) line is also almost 100 \kms\ wider than the $^{12}$CO(3-2) line observed here (see Figure \ref{spectrumplot} and \citealt{2011MNRAS.414..968D}). The $^{12}$CO(2-1) line was also observed by \cite{2011MNRAS.414..968D}, and although a slight pointing error means the spectrum is lopsided, it too appears to be wider than the $^{12}$CO(3-2) line.
Given the rising rotation curve revealed by Figure \ref{pvdplot}, this is naturally explained by a more radially extended low-$J$ CO distribution. 

\textsc{KinMS} modelling of the $^{12}$CO(1-0) data suggests that the surface brightness of the low-$J$ gas can be well fit by Equation \ref{eq:surfbrightprof}, but without the outer cutoff and plateau that was required for $^{12}$CO(3-2). The $^{12}$CO(1-0) is thus likely to have a similar inner profile shape, but with an uninterrupted exponential decrease towards the outer parts of the disc (echoing the findings of \citealt{2013MNRAS.429..534D}).

What causes this confinement of the hotter/denser $^{12}$CO(3-2) emitting gas to the inner regions? And what sets the inner and outer truncation radii?  $^{12}$CO(1-0) has a critical density of $\approx$1400 cm$^{-3}$ and an excitation temperature of 5.53 K, while $^{12}$CO(3-2) is excited in denser ($\approx7\times10^4$ cm$^{-3}$) and warmer ($\approx$15 K) gas.  In a well-mixed gas disc one would not expect a strong truncation of the hotter/denser gas, but a more gradual transition.
We investigate physical processes that can lead to this morphology in the following sections.

\subsubsection{Resonances}
\label{resonancesec}

Many of the ways to create rings and gaps in gas distributions rely on (rotating) non-axisymmetric features in the galaxy potential, that can lead to resonances throughout the disc (see e.g. the review by \citealt{1996FCPh...17...95B}).
 It is possible that the features we see in the gas distribution of NGC4429 are related to resonances due either to its large-scale bar (that may also have caused the outer {(pseudo-)ring} visible at large radii in Fig. \ref{gal_overview}), or a putative nuclear/secondary bar of a similar extent to the molecular gas disc itself. Indeed, the disc studied here is very reminiscent of the nuclear discs often observed within the inner Lindblad resonance (ILR) of barred disc galaxies. 

\begin{figure*} \begin{center}
\includegraphics[width=0.7\textwidth,angle=0,clip,trim=0cm 0cm 0cm 0.0cm]{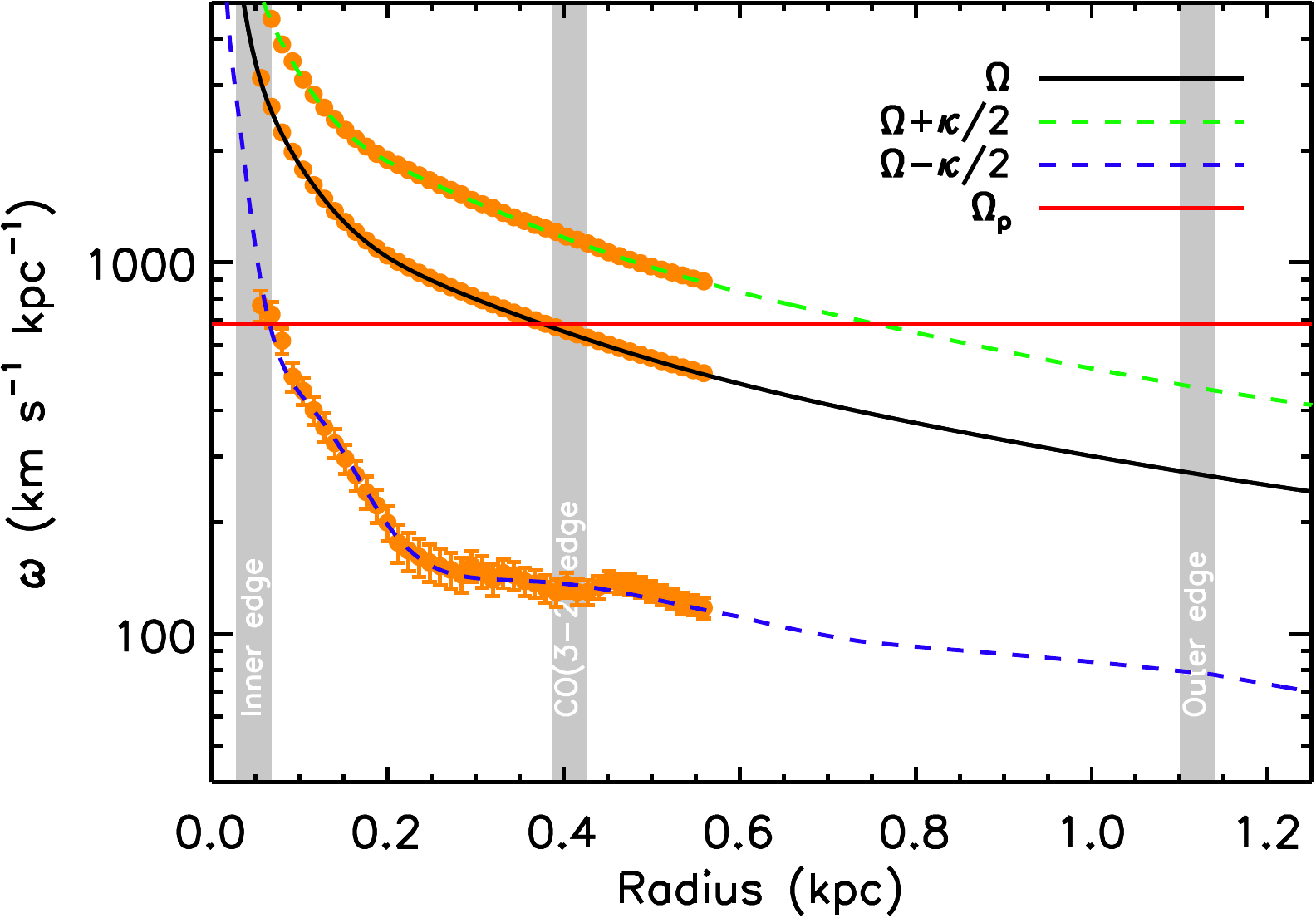}
\caption{Angular frequency (black solid line) and angular frequency plus (green dashed line) and minus (blue dashed line) the epicyclic frequency of the disc of NGC4429, extracted from our MGE models, as a function of radius. These curves define the location of the co-rotation and outer/inner Lindblad resonances, respectively. Also shown as orange data points with error bars are the same quantities extracted directly from our ALMA data. The grey bars show the location of the inner and outer truncation radii of the $^{12}$CO(3-2) disc, and the outer edge of the $^{12}$CO(1-0) disc. The red line shows a potential pattern speed leading to resonances at two of these locations. A single pattern speed can thus potentially explain some of the morphological features seen in the molecular gas disc of NGC4429, albeit with a very high pattern speed. }
\label{resonances}
 \end{center}
 \end{figure*}

We investigate this further in Figure \ref{resonances}. 
Let us define the angular frequency of a parcel of gas as $\Omega$ $\equiv |V_{\rm circ}|/R$, with $R$ the radius and $V_{\rm circ}$ the circular velocity at that radius. The epicyclic frequency $\kappa$, the frequency at which a gas particle oscillates 
about its guiding centre, is defined as

\begin{equation}
\kappa^2 =  4\Omega^2 + R\frac{d\Omega^2}{dR}.
\label{epicycle}
\end{equation}

\noindent Any perturbation to axisymmetry with a pattern speed $\Omega_{\rm p}$ has a co-rotation resonance at the point $\Omega = \Omega_{\rm p}$. The outer and inner Lindblad resonances are then found at $\Omega_{\rm p} = \Omega \pm \kappa/2$, respectively.

Here we use two different methods to estimate $\Omega$ and $\kappa$ for NGC4429.
Firstly we use the MGE model of the luminous mass, combined with the $M$/$L$ radial profile obtained from our best-fit \textsc{KinMS} model. We apply the $M$/$L$ gradient at $R<5$\arcsec\ only, and assume it flattens outside of this region, as seems to be required by the $^{12}$CO(1-0) data. We smooth the $M$/$L$ gradient between these two regimes to prevent an artificial break in $\kappa$ at $R=5$\arcsec, and include the best-fit SMBH as a point mass at the galaxy centre. The angular frequencies calculated in this manner are shown as black, green and blue lines in Figure \ref{resonances}. 

As a cross check we also calculate the same quantities directly from our data, by using the `trace' of the major-axis PVD, as defined in \cite{2013Natur.494..328D}. This trace allows us to estimate $V_{\rm circ}$ as a function of radius directly, from which we can calculate $\Omega$ and $\kappa$. These estimates of the angular frequencies are shown as orange data points in Figure \ref{resonances}, and agree well with our MGE determinations.

We do not know what pattern speed a nuclear bar or other asymmetry may have in NGC4429, but we can verify if any reasonable $\Omega_{\rm p}$ can lead to resonances at the locations of the breaks in the observed CO profiles. 
We show  the position of these breaks (the inner hole radius, the $^{12}$CO(3-2) maximum extent, and the $^{12}$CO(1-0) extent) as vertical grey bars in Figure \ref{resonances}. 
As the red line in Figure \ref{resonances} shows, a pattern speed of $\Omega_{\rm p} \approx 680$ km s$^{-1}$ kpc$^{-1}$ could cause resonances at the same locations as some of the structures we observe. The ILR would then lie at the inner edge of the disc and the $^{12}$CO(3-2) emission would extend out to the co-rotation radius only. In this model the $^{12}$CO(1-0) emission would extend much beyond the OLR. 

However, this pattern speed is very fast when compared to those of large-scale bars found in spiral galaxies, and for nuclear bars typically formed in simulations \citep[e.g.][]{1993A&A...277...27F,2002AJ....124...65E,2015A&A...576A.102A}. In addition, these simulations show that typically the large-scale and nuclear bars share resonances \citep{1990ApJ...363..391P}. We estimate that the large-scale bar visible in SDSS images (Figure \ref{gal_overview}) ends at a (de-projected) radius of $\approx$5.6 kpc, that would not correspond to any of the observed features. We note however that we do not know the behaviour of the stellar $M$/$L$ at these radii, and can only make the assumption that it stays constant. However, to force the co-rotation resonance of the nuclear bar to fall at the same location as the ILR of the primary bar (as is often the case in simulations; \citealt{1990ApJ...363..391P,2009ApJ...690..758S}) one would need extreme $M$/$L$ variation ($M$/$L_V$ $>$ 20~M$_{\odot}$/L$_{\odot,V}$ at $R$=\,5.6~kpc), which is likely unphysical. We thus consider a resonance origin for the gas morphology to be disfavoured, although we cannot rule it out entirely.

\subsubsection{Stability}
\label{stabilitysec}

The strong confinement of the hotter/denser $^{12}$CO(3-2) emitting gas to the inner regions of the molecular gas disc could naturally arise if the gas is more stable outside a radius of $\approx$450\,pc. If this were the case, gravitational perturbations that create dense gas and star formation within that radius could be suppressed outside of it. Support for this idea also comes from the \textit{HST} image presented in Figure \ref{gal_overview}. Inside this radius the dust distribution is interrupted by various bright clumpy structures, that are likely H{\small \textsc{II}} regions from ongoing star formation. These bright structures are less frequent outside this radius, suggesting star formation is less abundant in the outer disc. 

The \cite{1964ApJ...139.1217T} stability parameter for a gaseous disc is

\begin{equation}
Q =  \frac{\kappa \sigma_{\rm gas}}{\pi G \Sigma_{\rm gas}}\,,
\end{equation}
 
\noindent where $\kappa$ is the epicyclic frequency as defined in Eqn. \ref{epicycle}, $\sigma_{\rm gas}$ is the gas velocity dispersion, $G$ is the gravitational constant and $\Sigma_{\rm gas}$ is the surface density of the gas. 
 In a gas disc with $Q <1$, the self-gravity of the gas disc is sufficient to overcome the opposing 
 forces and lead to gravitational collapse, whereas $Q > 1$ indicates that the disc is supported and stabilised kinematically.
Note that while $Q > 1$ ensures stability against axisymmetric perturbations, larger values $(Q \approx2-3)$
are required to stabilise the disc against non-axisymmetric perturbations \citep{2017arXiv170102138R}.
Observations of spiral galaxies show that $Q$ typically spans the range $\approx$0.5\,--\,6 \citep[see e.g.][]{2001ApJ...555..301M}.

\begin{figure} \begin{center}
\includegraphics[width=0.49\textwidth,angle=0,clip,trim=0cm 0cm 0cm 0.0cm]{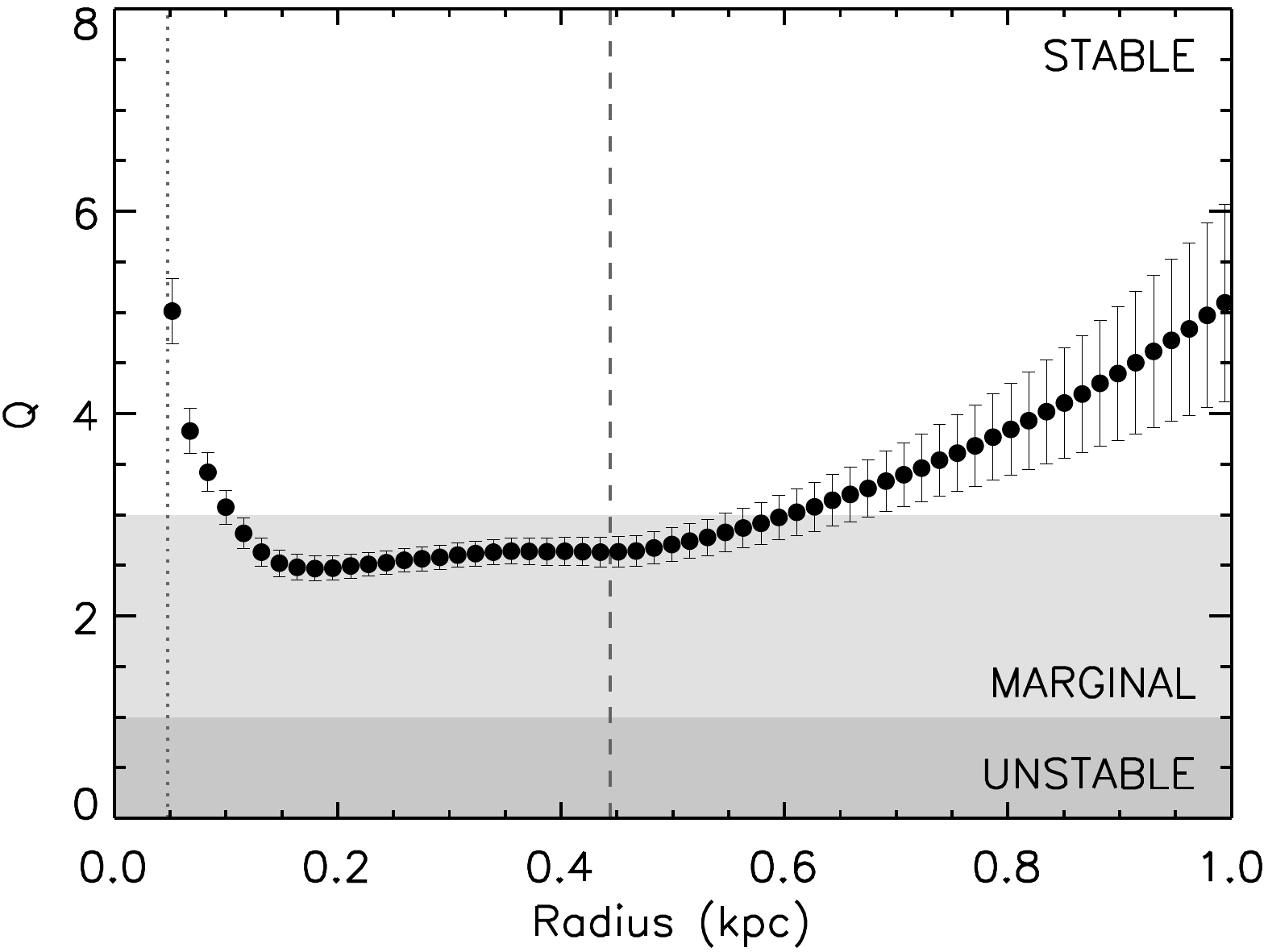}
\caption{Toomre Q parameter within the CO emitting disc of NGC4429. The dotted vertical line indicates the location of the central hole in the gas distribution. The dashed line indicates the edge of the $^{12}$CO(3-2) emitting disc, beyond which the disc is only traced in lower-$J$ transitions, likely because of the increasing stability in this region. }
\label{stability}
 \end{center}
 \end{figure}

Figure \ref{stability} shows the \cite{1964ApJ...139.1217T} stability parameter of NGC4429 as a function of radius, where we used the model of the total surface density profile $\Sigma_\mathrm{gas}(r)$ of the $^{12}$CO(1-0) emitting gas (as discussed in Section \ref{gasmorph}), normalised such that a thin disc with this profile has the same total molecular gas mass as found in \cite{2013MNRAS.432.1796A}. We fixed $\sigma_{\rm gas}$ to 2.2~\kms, as found in Section \ref{veldisp1}, and derived $\kappa$ as in Section \ref{resonancesec}.
We propagated the observational uncertainties on $\Sigma_\mathrm{gas}$ (2.5~M$_\odot$~pc$^{-2}$; \citealt{2013MNRAS.432.1796A}) and assumed an uncertainty of 10~\kms\ on $V_\mathrm{circ}$ (as above) to determine the uncertainties on $Q$.

Figure \ref{stability} shows that around the outer truncation of the dense $^{12}$CO(3-2)-emitting gas (indicated with a vertical dashed line), the disc of NGC4429 does appear to become more stable. 
Although it appears from this calculation that the entire disc is marginally stable, $Q$ is only an approximation and we did not include the effect of the stellar disc, which deepens the potential and makes the gas less stable  \citep[see e.g.][]{2011MNRAS.416.1191R}. 
To first order, however, the increased stability of the gas in the outer parts of the disc could help explain why the dense $^{12}$CO(3-2) emitting gas is confined to the inner parts of the galaxy (while the more tenuous CO(1-0) emitting gas is more extended).

The stability of the gas also increases in the galaxy centre. This increase starts at a radius of $\approx$125\,pc and reaches $Q$ $\approx$5 at $\approx$40\,pc.
\cite{2011PASP..123..514B} show that the dominant gas ionisation mechanism in NGC4429 changes at small radii. H{\small \sc II} regions dominate in the outer parts, while AGN photoionisation takes over at a radius of $\approx$120\,pc. This suggests the higher stability of the gas is affecting star formation in the inner parts of this galaxy. 
It is also possible that the increasing stability of the gas in the inner parts is linked to the formation of the hole in the gas disc. However, the radius of the hole is smaller than the radius at which the stability of the gas begins to increase. It is also possible that the hard radiation field around the low-luminosity AGN heats the gas in the region between $\approx$40 and 125\,pc, compensating for the lower density and allowing $^{12}$CO(3-2) emission to still be detected. High resolution $^{12}$CO(1-0) observations would be required to fully test this hypothesis.

\subsubsection{Tidal forces}

In the region around an SMBH the molecular clouds in galaxies are subject to strong tidal forces. As discussed by \cite{2005ApJ...628..169S}, tidal forces may play a role in suppressing star formation around the massive black holes of early-type disc galaxies, and thus could potentially help explain the central hole in the molecular cloud distribution of NGC4429.
Assuming the clouds are self-gravitating fluid bodies then the Roche limit, beyond which they break up, is given by

\begin{equation}
r_{\rm Roche} = 2.423\left(\frac{3M_{\rm BH}}{4 \pi \rho_{\rm cloud}}\right)^{1/3}\,,
\end{equation}
where $M_{\rm BH}$ is the SMBH mass and $\rho_{\rm cloud}$ is the mean density of the molecular cloud. 

Interestingly, this formula predicts that, for a SMBH mass of $1.5\times10^8$\,\msun, clouds with density less than $\approx9\times10^4$\,cm$^{-3}$ should not survive inside the radius of the observed molecular gas hole ($\approx$48\,pc). This density is very similar to the critical density of the $J$=3-2 CO transition we are observing here ($\approx9\times10^4$\,cm$^{-3}$). 
Of course denser clumps of gas could still survive inside this radius. The densest parts of molecular clouds reach $\approx10^7$\,cm$^{-3}$,  and such clouds could hold together against tidal forces until $\approx$10\,pc from the SMBH. 
Other forces (e.g. external pressure) can also provide additional confinement, and allow even low-density clouds to survive closer to the SMBH. Tidal forces alone thus cannot explain why we do not detect $^{12}$CO(3-2) in the very centre of NGC4429. 

It is, however, possible that a combination of tidal forces and other mechanisms mentioned above could be important. 
As non-circular motions in NGC4429 seem to be negligible, any cloud present has to have formed in this region.
The low gas velocity dispersion and increasing stability of the gas in the centre of NGC4429 (Section \ref{stabilitysec}) will tend to suppress cloud fragmentation and limit the density distribution of the clouds that do form. Hydrodynamic simulations suggest that in a stable low velocity dispersion disc, volume densities larger than $\approx10^3$\,cm$^{-3}$ are unlikely (e.g. left panel of Figure 4 in \citealt{2013MNRAS.432.1914M}).
If this were indeed the case, then within $\approx$50\,pc of the SMBH, where tidal forces are high, one would not expect new molecular clouds to form in NGC4429. 
However, in NGC4697, which has a very similar low dispersion molecular gas disc as NGC4429, gas is detected at very small radii around the SMBH. This argues that even if such a mechanism is important in NGC4429, it may not be important in all galaxies.
High resolution observations of other CO transitions, with different critical densities, will be required to determine the correct interpretation.

\subsection{Continuum emission}
\label{contdiscuss}

As discussed above, we only detect continuum emission from the nuclear region of NGC4429. This is despite the predicted intensity of 340 GHz thermal dust emission (from blackbody fits to the \textit{Herschel} photometry; \citealt{2013A&A...552A...8D}; Smith et al., in prep.) being significantly higher than our detection threshold. It is thus likely that we are resolving out emission from the large-scale dust disc, and are left with only emission from the central regions of the galaxy. As such are unable to study the dust properties of the clouds in the main disc of the galaxy. 

One can, however, consider what mechanism is powering the emission we do detect from the central regions of NGC4429. For the central source, our measured 340 GHz (883\,$\mu$m) integrated flux density of 1.38\,$\pm$\,0.07\,$\pm$0.14 mJy is higher than the 5 GHz flux density (0.40\,$\pm$\,0.02 mJy) reported by \cite{2016MNRAS.458.2221N}. These two measurements yield a radio spectral index of 0.29$\pm$0.03. At these high frequencies both synchrotron and free-free emission are expected to have negative spectral indices.
In addition, CARMA observations from \cite{2013MNRAS.432.1796A} set an upper limit to the continuum flux at 115 GHz of 0.375 mJy, effectively ruling out synchrotron and free-free radiation as the source of the detected emission.

We thus suggest that we are detecting thermal dust emission from material in the torus and narrow-line region heated by a low-luminosity AGN. The fact that this emission is marginally resolved suggests there must be a dust structure at the centre of NGC4429, that is perhaps too hot to be traced well by mid-$J$ molecular gas tracers.

 \subsection{Gas velocity dispersion}
 \label{veldispdiscuss}

As discussed above, the velocity dispersion in the molecular gas disc of NGC4429 seems abnormally low. 
A similar signature was also seen in NGC4697 \citep{2017arXiv170305248D}. In that object part of the explanation seemed to be that the total molecular gas mass was low, making it unlikely that the GMC mass function was fully populated. The GMC population of NGC4429 will be studied in detail in a future work (Liu et al., in prep.).  However, in this source, the total molecular mass is an order of magnitude higher than in NGC4697, suggesting that the GMC mass function will be much better sampled. 

We are thus left searching for additional physical mechanisms that can explain this low velocity dispersion. 
As discussed in \cite{2017arXiv170305248D}, one possible explanation of this discrepancy is a hard radiation field, that may indeed be present in NGC4429 (that is embedded deep in the hot X-ray halo of the Virgo cluster and has a low-luminosity AGN). 
Given the $\approx$15 pc resolution of our observations, each line-of-sight likely intersects only a single cloud, and the dispersion we measure is likely dominated by intra-cloud motions. In a hard radiation field, it is possible that the CO molecules are confined deeper inside the molecular clouds, in dynamically colder regions where both thermal and non-thermal motions would be reduced \citep{2011MNRAS.415.3253S,2015MNRAS.452.2057C}.  

Evidence has been mounting in recent years that cloud-scale gas structures do vary within and across galaxies \citep[e.g.][]{2005PASP..117.1403R,2013ApJ...779...46H,2015ApJ...801...25L,2015ApJ...803...16U}. 
\cite{2016ApJ...831...16L} have shown that the gas velocity dispersion of molecular clouds correlates strongly with their surface density (their Equation 13). The measured gas velocity dispersion of NGC4429, 2.2$^{+0.68}_{-0.65}$~\kms, suggests that the mean surface density of its molecular clouds is around 2.6 \msun\,pc$^{-2}$ (if this relation holds in sources of this type).
Assuming a Galactic CO-to-H$_2$ conversion factor and the measured CO(3-2)/CO(1-0) ratio of 1.06 we find that the observed molecular cloud surface densities in NGC4429 are significantly higher than this, varying from $\approx$40 to $\approx$1000 \msun\,pc$^{-2}$ in $\approx$60$\times$60 pc$^2$ resolution elements. We thus conclude that the surface densities of these clouds are unlikely to be the reason they have low velocity dispersions. We will explore the discrepant behaviour of the clouds in this object in detail in a future work (Liu et al., in prep.).

Another possible explanation for the low gas velocity dispersion is the stabilising influence of the galaxy bulge (so-called `morphological quenching'). \cite{2009ApJ...707..250M,2013MNRAS.432.1914M} showed using hydrodynamic simulations that the presence of a large bulge can stabilise a low-mass gas disc against star formation, leading to a low velocity dispersion. In Section \ref{stabilitysec} we found that the gas disc of NGC4429 is marginally stable, as would be expected in this scenario. If `morphological quenching' were playing a role in this object, however, we would also expect it to have a low star formation efficiency (SFE). But, unlike NGC4697, NGC4429 appears to have a fairly normal SFE \citep{2014MNRAS.444.3427D}, making this explanation less appealing. 

  Further observations of galaxies with dynamically cold gas discs are required to confirm which, if any, of these mechanisms can explain this phenomenon.

\section{\uppercase{Conclusions}}
\label{conclude}

In this work we presented observations, taken as part of the WISDOM project, of the nearby fast-rotating early-type galaxy NGC4429, that has a boxy/peanut-shaped bulge. We observed $^{12}$CO(3--2) emission from this object with the Atacama Large Millimeter/submillimeter Array (ALMA) in cycle-2, with a linear resolution of 14$\times$11 pc$^2$ (0\farc18\,$\times$\,0\farc14). 

NGC4429 has a flocculent molecular gas disc of radius $\approx$400\,pc, with a central hole of radius $\approx$40\,pc. 
The warm/dense $^{12}$CO(3-2) emitting gas appears to be confined to the inner parts of an obscuring dust disc visible in \textit{HST} images, possibly as a consequence of the increasing stability of the gas in the disc outer regions. Suppressed fragmentation and high tidal shear may act together to prevent the formation of molecular clouds close to the very centre of this galaxy, creating the observed molecular hole.

A forward modelling approach in a Bayesian framework was used to fit the observed data cube of NG4429 and estimate the SMBH mass, stellar $M$/$L$, and numerous parameters describing the structure of the molecular gas disc.
A strong $M$/$L$ gradient is required in the central 200 pc, although this gradient includes both real $M$/$L$ variations and the effects of potential errors in the mass modelling.

We estimated a mass of  (1.5$\pm0.1^{+0.15}_{-0.35}$)$\times$10$^8$ \msun\ for the SMBH in the centre of NGC4429, where the two uncertainties reflect the random and systematic uncertainties, respectively.
This SMBH mass is fully consistent with the previous upper limit of 1.8~$\times$~10$^8$~\msun\ set by \cite{2009ApJ...692..856B}.
NGC4429 lies slightly above the best-fit $M_{\rm BH}$ -- $\sigma_{\rm e}$ relation of \cite{2016ApJ...831..134V}, but well within the scatter.

NGC4429 was found to have a very low molecular gas velocity dispersion, similar to that found in NGC4697 \citep{2017arXiv170305248D}, but in this object the GMC mass function is likely to be well sampled, so another physical mechanism is likely required to explain the low dispersion. 
It is possible that the large bulge of NGC4429 stabilises the gas disc (as expected from simulations of morphological quenching), but the SFE appears normal, and hence this explanation is disfavoured. 
We suggest that the CO molecules may be confined deeper inside the molecular clouds by a hard cluster radiation field, but clearly more data are required to truly understand the origin of these incredibly low velocity dispersions.

This SMBH measurement using molecular gas dynamics, the sixth presented in the literature, once again demonstrates the power of ALMA to constrain SMBH masses. As the technique comes of age, it will become possible to probe secondary black hole -- galaxy correlations, and shed further light on the physics behind the co-evolution of SMBHs and their host galaxies.

 \vspace{0.5cm}
\noindent \textbf{Acknowledgments}

The authors thank the referee for comments that improved this paper.
TAD acknowledges support from a Science and Technology Facilities Council Ernest Rutherford Fellowship, and thanks K. Nyland, P. Clark, A. Whitworth and N. Peretto for useful discussions. 
MB was supported by the consolidated grants `Astrophysics at Oxford' ST/N000919/1 and ST/K00106X/1 from the UK Research Council. FvdV acknowledges support from the Klaus Tschira Foundation.
MC acknowledges support from a Royal Society University Research Fellowship.

This paper makes use of the following ALMA data: ADS/JAO.ALMA\#2013.1.00493.S. ALMA is a partnership of ESO (representing its member states), NSF (USA) and NINS (Japan), together with NRC (Canada) and NSC and ASIAA (Taiwan) and KASI (Republic of Korea), in cooperation with the Republic of Chile. The Joint ALMA Observatory is operated by ESO, AUI/NRAO and NAOJ.
This paper also makes use of observations made with the NASA/ESA Hubble Space Telescope, and obtained from the Hubble Legacy Archive, which is a collaboration between the Space Telescope Science Institute (STScI/NASA), the Space Telescope European Coordinating Facility (ST-ECF/ESA) and the Canadian Astronomy Data Centre (CADC/NRC/CSA). This research has made use of the NASA/IPAC Extragalactic Database (NED) which is operated by the Jet Propulsion Laboratory, California Institute of Technology, under contract with the National Aeronautics and Space Administration.

\bsp
\bibliographystyle{mnras}
\bibliography{bibNGC4429.bib}
\bibdata{bibNGC4429.bib}
\bibstyle{mnras}

\label{lastpage}

\end{document}